\documentclass[a4paper,fleqn,usenatbib]{mnras}

\usepackage{newtxtext,newtxmath}

\usepackage[T1]{fontenc}
\usepackage{ae,aecompl}
\usepackage{tablefootnote}

\usepackage{graphicx}	
\usepackage{amsmath}	
\usepackage{amssymb}	
\usepackage{url}

\usepackage{etoolbox}
\makeatletter
\patchcmd\@combinedblfloats{\box\@outputbox}{\unvbox\@outputbox}{}{\errmessage{\noexpand patch failed}}
\makeatother
\def\bl{Blazhko}
\def\fz{$f_{0}$}
\def\fm{$f_{\rm m}$}

\def\pp{$P_{\rm Puls}$}
\def\pb{$P_{\rm BL}$}

\title[Galactic bulge Blazhko RR Lyrae stars II]{Blazhko effect in the Galactic bulge fundamental mode RR~Lyrae stars II: Modulation shapes, amplitudes and periods}

\author[Skarka,\,Prudil\,\&\,Jurcsik]{
M. Skarka$^{1,2}$\thanks{marek.skarka@asu.cas.cz}, Z. Prudil$^{3}$ \& J. Jurcsik$^{4}$ 
\\
$^{1}$ Department of Theoretical Physics and Astrophycis, Masaryk University, Kotl\'{a}\v{r}sk\'{a} 2, 611 37 Brno, Czech Republic\\
$^{2}$ Astronomical Institue of the Czech Academy of Sciences, Fri\v{c}ova 298, 251 65 Ond\v{r}ejov, Czech Republic\\
$^{3}$ Astronomisches Rechen-Institut, Zentrum f{\"u}r Astronomie der Universit{\"a}t at Heidelberg, M{\"o}nchhofstr. 12-14, 69120 Heidelberg, Germany\\
$^{4}$ Konkoly Observatory, H-1121 Budapest, Konkoly Thege Mikl\'{o}s  \'{u}t 15-17., Hungary
}

\date{Accepted XXX. Received YYY; in original form ZZZ}

\pubyear{2016}

\begin{document}
\label{firstpage}
\pagerange{\pageref{firstpage}--\pageref{lastpage}}
\maketitle

\begin{abstract}
The number of stars observed by the Optical Gravitational Lensing Experiment (OGLE) project in the Galactic bulge offers an invaluable chance to study RR~Lyrae stars in a statistical manner. We used data of 3141 fundamental-mode RR~Lyrae stars showing the Blazhko effect observed in OGLE-IV to investigate a possible connection between modulation amplitudes and periods, light curve and pulsation characteristics. We found that there is no simple monotonic correlation between any combination of two parameters concerning the Blazhko and pulsation amplitudes, periods and the shape of the light curves. There are only systematic limits. There is a bottom limit of the modulation period with respect to the pulsation period. We also found that the possible range of modulation amplitudes decreases with increasing pulsation period which could point towards that the Blazhko effect is suppressed in cooler, larger, more luminous and less metal abundant bulge RR~Lyrae stars. Our investigation revealed that the distribution of the modulation periods can be described with two populations of stars with the mean modulation periods of 48 and 186\,days. There is a certain region with a low density of the modulated stars, which we call the Blazhko valley, in the pulsation period-modulation period plane. Based on the similarity of the modulation envelopes, basically every star can be assigned to one of six morphological classes. Double modulation was found in 25~per~cent of the studied stars. Only 6.3~per~cent of modulated stars belongs to the Oosterhoff group II. 
\end{abstract}


\section{Introduction}\label{Sect:Introduction}

Various studies show that about 40-50~per cent of classical RR Lyrae stars (hereafter RRLs) pulsating in the fundametal mode show the quasi-periodic variations in amplitude and phase of light curves \citep[see][]{kovacs2016}. \citet{kovacs2018} even proposes based on {\it K2} data that up to 100~per cent of RRLs has the \bl~effect, which is, however, in contrast to results by \citet{Plachy2019} who used the method of extended aperture photometry and detected only 45~per cent of stars to be modulated. This light curve modulation, which is usually called the Blazhko effect \citep{blazhko1907}, lacks the explanation even a century after its discovery.

Most of the proposed models face problems to explain one or more observed characteristics \citep[see the discussion in][]{kollath2018}. Currently, it seems that the most likely explanation of the modulation is in the resonance 9:2 between the fundamental and the ninth overtone modes \citep{kollath2011,kollath2018}. The same resonance can explain the period doubling observed in \bl~stars \citep{buchler2011}. This model is in accordance with the current findings by \citet{jurcsik2017,jurcsik2018a,jurcsik2018b}, who revealed that radius changes connected to the modulation occur only in the outer-most layers of the atmosphere, and the change of the radius at the photospheric layers is constant during the modulation cycle. Consequently,  the modulation is caused primarily by changes in the temperature variation.

Large sample(s) of stars are needed to fully cover the stunning diversity of the modulation properties \citep[see e.g.][]{benko2010}. A successful model must comply with describing the large variety in amplitudes, periods and shape of the modulation. Even with a good model of the modulation, there still remains lots of questions to be answered. Why some stars show modulation and others not when there is no difference in the mean characteristics of the modulated and non-\bl~stars \citep{smolec2005,skarka2014b,prudil2017}? Is the occurrence of the modulation driven by some particular physical property \citep[e.g. temperature, chemical composition,][]{jurcsik2011,arellano2016a}? Why is the occurrence of modulated RRLs lower among stars with longer pulsation periods \citep{jurcsik2011,skarka2014b}? Is there any relation between the modulation period and light curve characteristics \citep{jurcsik2005a,jurcsik2005b,benko2014a}? 

The aim of our study is to describe the modulation and investigate the possible relations between modulation and pulsation light curve characteristics. Thus, we investigate the relations between modulation periods, amplitudes, shapes of the light curves and frequency spectra. We also describe the modulation envelopes and define six basic morphological types of the modulation shapes. At last, we investigate the possible connection between the \bl~effect and the Oosterhoff dichotomy \citep{oosterhoff1939}. We do not deal with detailed investigation of the frequency spectra of particular stars and period variations at all.

The whole study is based on the paper by \citet{prudil2017}, where we carefully analysed OGLE-IV data \citep{soszynski2014} of more than 8000 fundamental-mode RRLs located in the Galactic bulge. We identified more than 3000 Blazhko stars, which creates the most extensive sample of modulated RRLs ever studied. Because the data was gathered with one telescope with the same detector, with similar cadence and time span, this gives a unique chance to get a reliable overall picture of the modulation periods and amplitudes among the Blazhko stars.

The paper is structured as follows: In Sect.~\ref{Sect:Sample}, we briefly describe the sample of the stars, give details about methods and definition of the parameters we use, and define the six modulation classes. Results on modulation properties can be found in Sect.~\ref{Sect:Results}. We discuss our results in Sect.~\ref{Sect:Discussion} and sum up the content of the paper in Sect.~\ref{Sect:Summary}.

\section{The sample and methods}\label{Sect:Sample}

In \citet{prudil2017}, we found 3341 stars to show the \bl~effect based on the look of the frequency spectra and the presence of the equidistant side peaks with amplitudes above {\it SNR\footnote{Signal-to-Noise Ratio}}~$>3.5$. This sample was studied again employing automatic procedures and applying subsequent visual inspection of all the frequency spectra of the studied stars to discard the stars with ambiguous modulation. Since we are now interested in the characteristics of the Blazhko effect, we performed additional data examination and analysis. 

\subsection{Frequency spectra}\label{Subsect:FreqSpec}

The modulation period $P_{\rm BL}$ can be defined as the reciprocal value of the difference between the side peaks and the basic pulsation frequency in the frequency spectrum. As the value of the modulation frequency \fm~we chose the one defined by the side peak with higher {\it SNR}: 
\begin{equation}\label{Eq:BlazhkoPeriod}
    P_{\rm BL}=1/|f^{\rm SNR}_{0\pm}-f_{0}|=1/f_{\rm m},
\end{equation} 
where $f^{\rm SNR}_{0\pm}$ stands for the side peak with the highest {\it SNR} close to $f_{0}$. From our previous analysis performed in \citet{prudil2017}, we roughly knew the length of the modulation period, which allowed us to automatically identify the position of the side peaks in all sample stars. 

First we fit the data with 10 harmonics using the \texttt{lcfit} routine \citep{sodor2012a} to refine the pulsation frequency $f_{0}$ and its uncertainty\footnote{The uncertainty of $f_{0}$ is the formal error resulting from the least-squares method.} (columns 2 and 3 in Table~\ref{Tab:Parameters}) to estimate and the mean magnitude $I_{\rm mean}$ and the mean amplitude $A_{\rm mean}$ (columns 13 and 16 in Table~\ref{Tab:Parameters}).The mean magnitude $I_{\rm mean}$ is the zero point of the fit. The mean amplitude $A_{\rm mean}$ is simply defined as the difference between two points with maximum and minimum brightness of the 10-harmonics model with 1000 points over the pulsation cycle. Subsequently, we pre-whitened the frequency spectra with 10 pulsation harmonics keeping the basic pulsation frequency fixed and searched for the highest peaks within $\pm0.0005$\,c/d around the expected positions $f_{0}\pm f_{\rm m}$. This interval was found as optimal for the automatic proper identification of the peaks. The value 0.0005\,c/d roughly corresponds to the inverse of the time span of the data for most of the stars. 

In case of positive identification including visual check we extracted the exact position of the peaks ($f_{+}=f_{0}+f_{\rm m}$ and $f_{-}=f_{0}-f_{\rm m}$ for the right-hand and left peak, respectively). Then we re-fit the full data set with 10 harmonics of $f_{0}$ (fixed) including $f_{+}$ and $f_{-}$ which were left as free parameters. We did not include any other frequency since this would need careful and deep frequency analysis of every data set which is out of scope of this paper. After this fitting procedure, we got $f_{-}$ and $f_{+}$ (columns 7 and 10 in Table~\ref{Tab:Parameters}) including their amplitudes $A_{+}$, $A_{-}$ (columns 9 and 12 in Table~\ref{Tab:Parameters}) and we were able to calculate the Blazhko period $P_{\rm BL}$ (Eq.~\ref{Eq:BlazhkoPeriod} and its error using the Gauss law of error propagation, columns 5 and 6 in Table~\ref{Tab:Parameters}). We also got information about the amplitude of the residuals of the frequency spectra in the $\pm$0.3\,c/d vicinity of $f_{0}$ after removing 10 pulsation frequency harmonics defining the noise level ({\it ARES$_{f_{0}\pm0.3}$}, column 4 in Table~\ref{Tab:Parameters}). 

The knowledge of the position of the side peaks and their amplitudes allowed us to compute the amplitude asymmetry parameter $Q$ defined by \citet{alcock2003} as 
\begin{equation}\label{Eq:Q}
Q=\frac{A_{+}-A_{-}}{A_{+}+A_{-}},
\end{equation}
and the frequency asymmetry parameter 
\begin{equation}\label{Eq:df}
\Delta f=f_{+}+f_{-}-2f_{0}.
\end{equation}

\begin{figure}
    \includegraphics[width=\columnwidth]{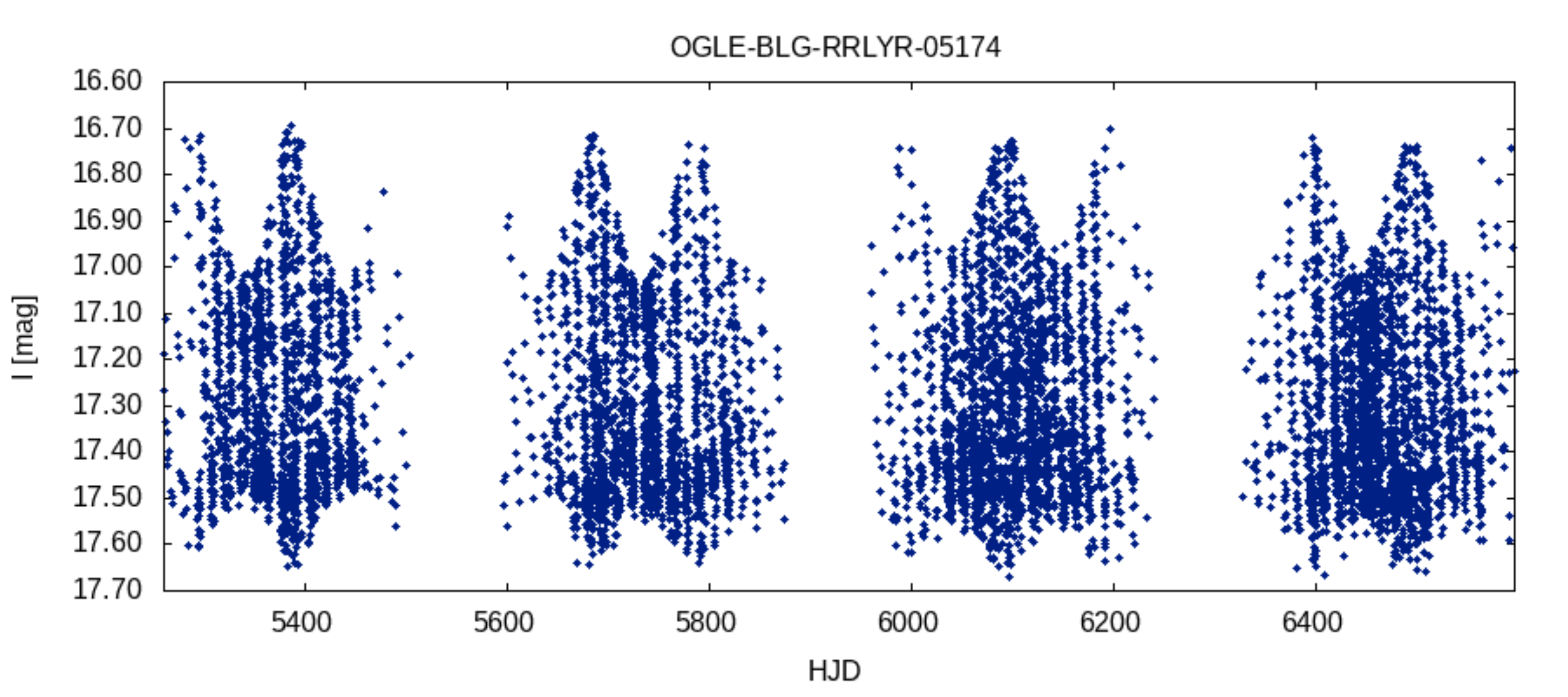}\\
	\includegraphics[width=\columnwidth]{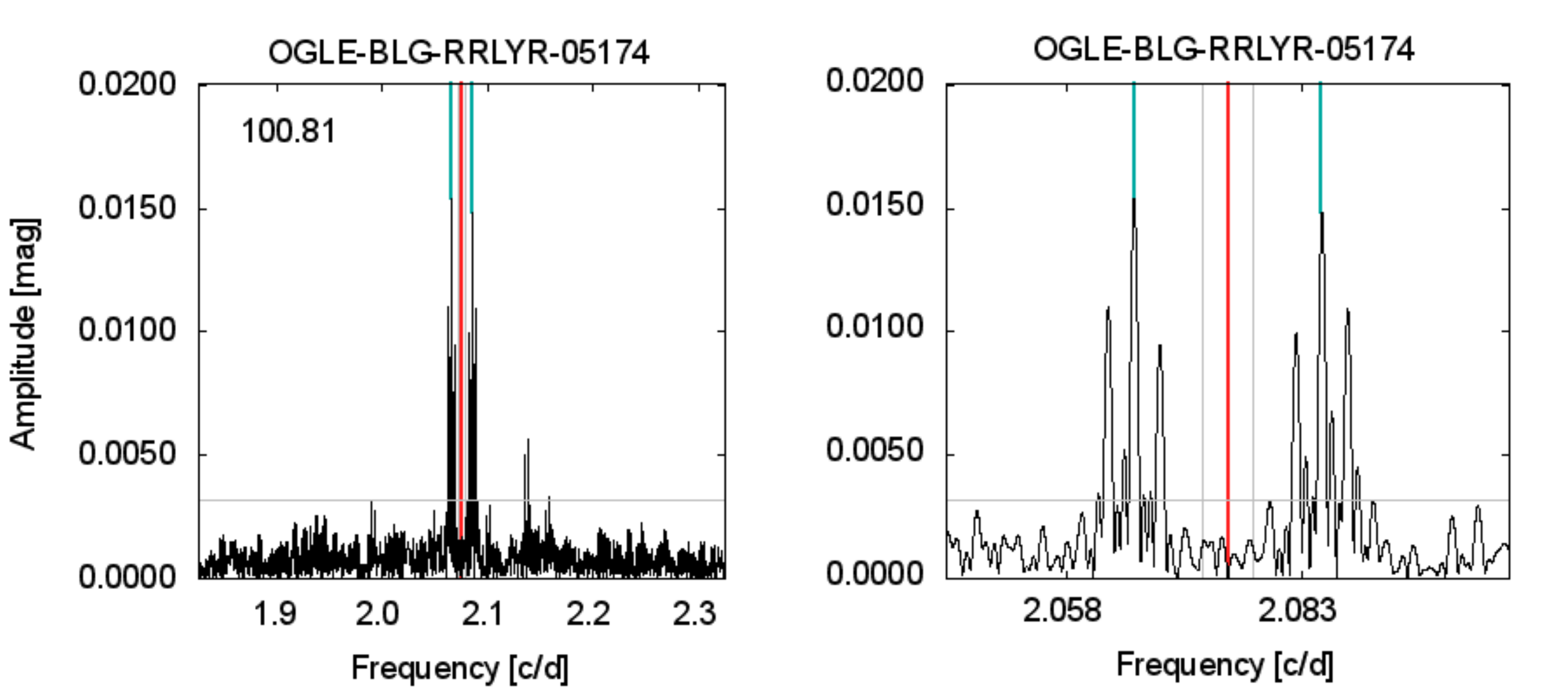}\\
	\includegraphics[width=\columnwidth]{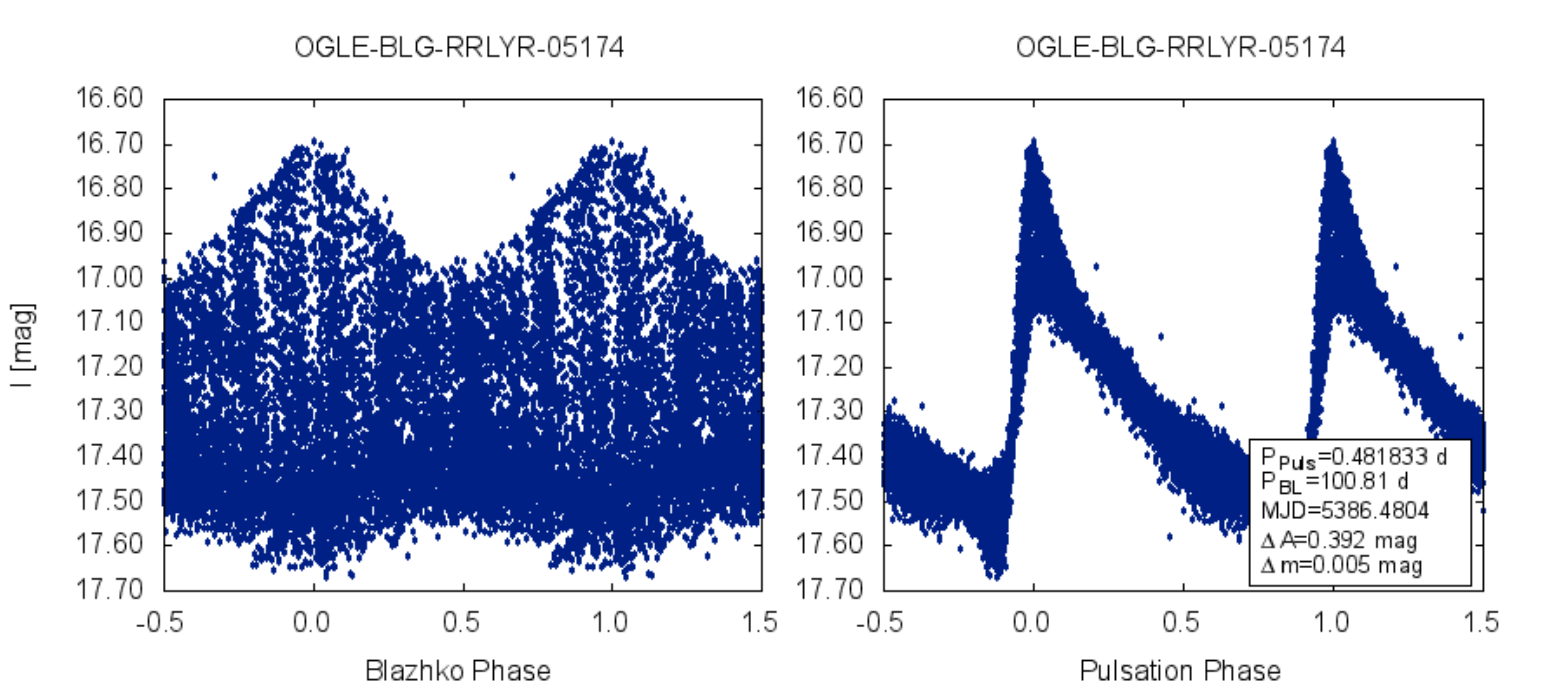}
    \caption{Examples of plots that we used for visualisation while studying the sample stars: the data ({\it top panel}), frequency spectra ({\it the two middle panels}) and Blazhko- and pulsation-phased light curves ({\it the two bottom panels}). The vicinity of the basic pulsation frequency (red continuous line) is zoomed for better readability in the middle right-hand panel. The vertical cyan lines show the exact positions of the side peaks, the vertical grey lines show the positions of yearly aliases of \fz, and finally, the horizontal grey line shows the {\it SNR}~$\sim3.5$~limit. The bottom right-hand panel shows the data phased with the modulation period, while the bottom left panel shows the folded light curve with the basic pulsation frequency. Additional information about the modulation characteristics is given in the bottom right corner of the bottom right-hand panel.}
    \label{Fig:FreqPhase}
\end{figure}

For each of the sample stars, we plotted the data, the frequency spectra for a fixed $\pm0.3$\,c/d vicinity of the basic pulsation frequency and a plot showing more details for a smaller range defined by the the modulation frequency (Fig.~\ref{Fig:FreqPhase}).   

For the stars that have longer modulation period than the half of the time span of the OGLE-IV data (till the end of 2014)\footnote{In 2018 the new data up to 2017 were released. We did not use these data.} we stitched the data with OGLE-III data to get longer time base \citep[for details see sect. 3 in][]{prudil2017}. Then, the frequency spectra and phase light curves of all stars were visually inspected. We discard stars that have too scattered data, showed bunch of peaks next to the basic pulsation frequency or modulation frequencies signalizing some long-term variations. We also removed stars that appeared to show side peaks with {\it SNR} below 3.5 and stars with modulation period longer than 2/3 of the longest possible time base (OGLE-III + OGLE-IV).

After this procedure we were left with 3141 stars. We include information about additional peaks in the vicinity of the basic pulsation frequency that could signalize additional modulation (Additional Peaks -- AdP) or long-term period change (Unresolved peaks -- UP). The full table is available only electronically as a supplementary material to this paper and the example of a few first rows is shown in Table \ref{Tab:Parameters}.

\subsection{Amplitudes, rise time and envelope parameters}\label{Sect:AmplitudeDetermination}

To get the amplitude of the modulation and variation of the mean magnitude, we first phased the data with the modulation frequency with the zero epoch $t_{0}$ corresponding to the brightest point in the data set. In vast majority of the stars, the brightest point corresponds to the maximal amplitude during the Blazhko effect. Then we separated the data into 10 or 20 bins centered to $t_{0}$ taking into account the number of points\footnote{Ten bins for stars with less than 1000 points, 20 bins for stars with more data points.} and modelled the light curve in each bin $j$ with the Fourier series
\begin{equation}\label{Eq:FourSum}
m_{j}(t)=m_{0,j}+\sum^{N_{j}}_{i=1}a_{i,j}\sin\left(2\upi i\frac{t-t_{0}}{P_{\rm puls}}+\varphi_{i,j}\right),
\end{equation}
where $t_{0}$ is the zero epoch, $a_{i,j}$ and $\varphi_{i,j}$ are the amplitudes and phases of the $i$-th component in $j$-th bin, respectively. The degree of the fit in the $j$-th bin $N_{j}$ was selected in such a way that the amplitude of the $N_{j}$-th component was four times higher than the noise level. Typically, $N_{j}$ was between 4 and 10.

The total amplitude $A_{j}$ in a particular \bl~phase (in the $j$-th bin) was defined simply as the difference between the brightest and faintest point of the fit in the same way as the value of the mean amplitude $A_{\rm mean}$ -- as the difference between two model points with maximum and minimum brightness. The amplitude of the modulation is defined in two comparable ways. First as the difference between the maximal $A^{\rm MAX}_{\rm mean}$ and minimal $A^{\rm MIN}_{\rm mean}$ photometric pulsation amplitudes (columns 18 and 17 in Table~\ref{Tab:Parameters})
\begin{equation}\label{Eq:AblPuls}
A^{\rm BL}_{\rm Puls}=A^{\rm MAX}_{\rm mean}-A^{\rm MIN}_{\rm mean}.
\end{equation}
The second definition of the modulation amplitude we adopted is the sum of the top and bottom amplitude of the envelopes ($A^{\rm TOP}_{\rm env}$ and $A^{\rm BOT}_{\rm env}$; columns 19 and 20 in Table~\ref{Tab:Parameters}): 
\begin{equation}\label{Eq:AblEnv}
A^{\rm BL}_{\rm env}=A^{\rm TOP}_{\rm env}+A^{\rm BOT}_{\rm env}.
\end{equation}
To get this information, We divided the Blazhko-phased data into 20 bins and got the highest and lowest points of the envelopes in each bin. With such definitions, the zero epoch of the modulation plays no role, because the maximal and minimal amplitudes are chosen regardless on the original Blazhko phasing described above.

The amplitude of the variation of the mean magnitude during the Blazhko cycle is defined as
\begin{equation}\label{Eq:MeanVariation}
\Delta I=I^{\rm MAX}_{\rm mean}-I^{\rm MIN}_{\rm mean},
\end{equation} 
where $I^{\rm MAX}_{\rm mean}$ and $I^{\rm MIN}_{\rm mean}$ are the mean magnitudes during maximal and minimal amplitudes of the Blazhko cycle, respectively (columns 14 and 15 in Table~\ref{Tab:Parameters}). As such, they do not necessarily correspond to the maximal mean-magnitude difference during the cycle.

From the binning of the Blazhko-phased light curve we also have information about the rise times ({\it RT$_{\rm env}$}) of the envelopes, which correspond to the phase differences between maximum and minimum brightness of the envelopes:
\begin{equation}\label{Eq:RT}
    RT_{\rm env}=\phi^{\rm MAX}_{\rm env}-\phi^{\rm MIN}_{\rm env}.
\end{equation} 
The exact minimal and maximal phases $\phi^{\rm MIN}_{\rm env}$ and $\phi^{\rm MAX}_{\rm env}$ of the envelopes were determined as the minimal and maximal values of the fit of the 20 points from the 20 bins. The low-degree Fourier series employing the Blazhko period was the model function of the fit. Through the text, we mention exclusively the {\it RT$^{\rm TOP}_{\rm env}$} (column 21 in Table~\ref{Tab:Parameters}) of the top envelope since it is better defined and is very similar to the bottom envelope in most stars.

\subsection{Morphological types}\label{Sect:MorphologicalTypes}

Based on the visual inspection of the Blazhko-phased light curves, we divided the star into six basic morphological types:
\begin{figure*}
    \includegraphics[width=2\columnwidth]{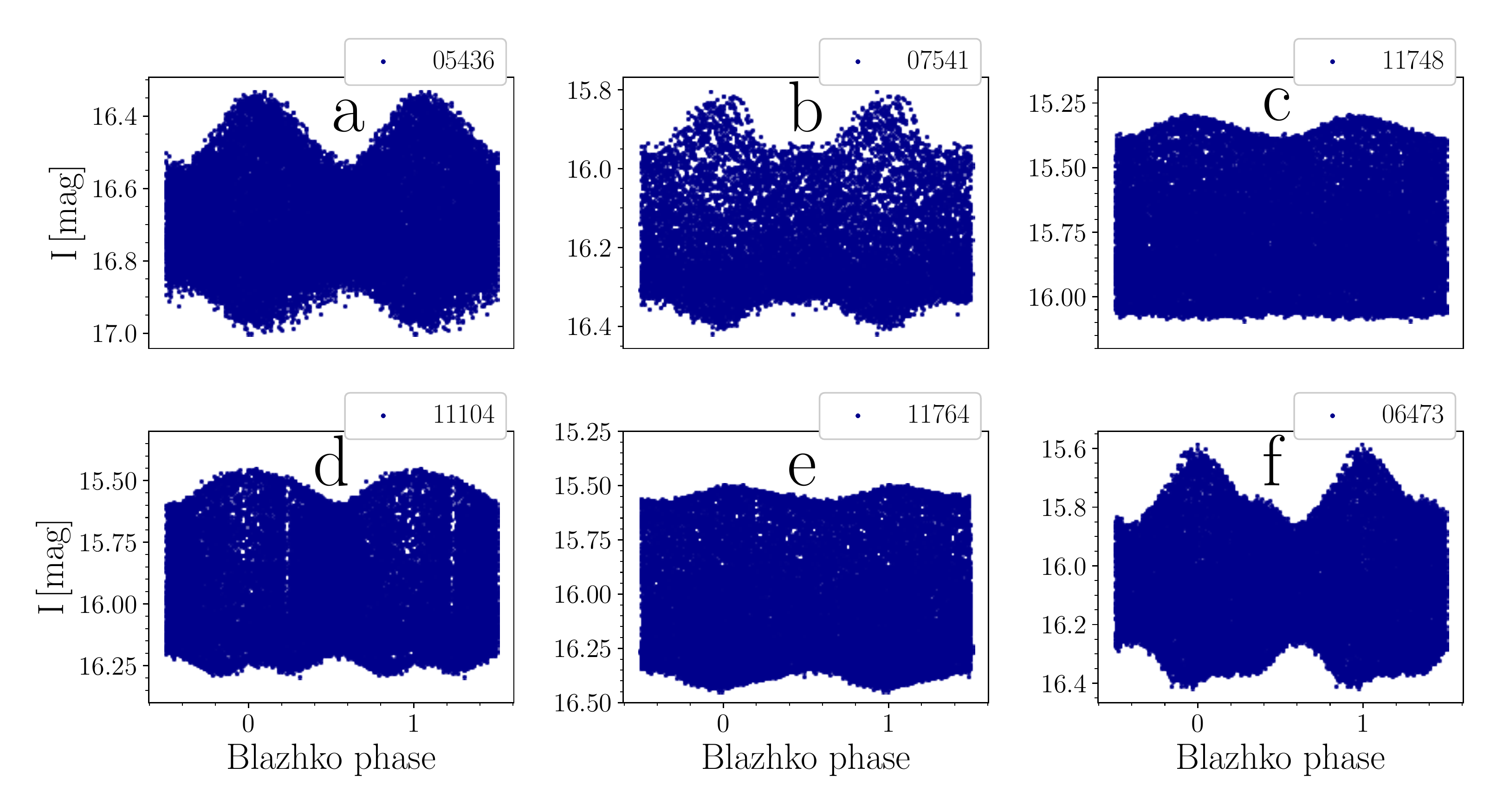}
    \caption{The prototypes of the basic morphological types showing different shape of the modulation envelopes. The data are phased with the Blazhko period, the legend of each panel denotes the ID of a given star in the form OGLE-BLG-RRLYR-ID. See the text for more details.}
    \label{Fig:ModulationTypes}
\end{figure*} 
\begin{itemize}
    \item {\it Class a} -- The top and bottom envelopes are nearly symmetric and sinusoidal and are close to anti-phase (the top-left panel of Fig.~\ref{Fig:ModulationTypes}). Sometimes the top envelope has a sharp maximum.
    \item {\it Class b} -- The Blazhko phase curve is flat during minimal modulation amplitude and the phase modulation well apparent (the top-middle panel of Fig.~\ref{Fig:ModulationTypes}).
    \item {\it Class c} -- The bottom envelope is flat (the top-right panel of Fig.~\ref{Fig:ModulationTypes}).
    \item {\it Class d} -- The bottom envelope has double minimum (the bottom-left panel of Fig.~\ref{Fig:ModulationTypes}).
    \item {\it Class e} -- The bottom envelope has larger amplitude than the top envelope (the bottom-middle panel of Fig.~\ref{Fig:ModulationTypes}).
    \item {\it Class f} -- The top and/or bottom envelope has a bump (the bottom-right panel of Fig.~\ref{Fig:ModulationTypes}).
\end{itemize}

It is possible that some morphological types are contaminated with RRLs from other classes because the sorting is based on a pure visual inspection. In addition, it is also possible that some of the types actually constitute only one class (see Sect.~\ref{Sect:Discussion}). We did not assign stars with low number of points and/or with small amplitude and/or bad phase coverage with any of the thypological classes. Only 2449 stars (78\,per cent) were assigned to one of the types. The basic statistical properties of the various classes can be found in Table~\ref{Tab:MorphTypes}.

\begin{table*}
\caption{The basic statistical data of the stars which were assigned to one of the six classes. N in the second column is the number of stars in a given sample, Rate shows the percentage of the stars according to the full sample, $\left<P_{\rm Puls}\right>$ is the mean pulsation period, $\left<A_{\rm Puls}\right>$ is the mean pulsation amplitude, $\left<P_{\rm BL}\right>$ is the mean Blazhko period, while $\left<A_{\rm BL}\right>$ shows the mean modulation amplitude. The tilde over variables stands for median value. The last column shows the percentage of the Oosterhoff\,II stars in a given sample.}
	\begin{tabular}{ccccccccccccc} 
		\hline
	ID & N & Rate & 
	     $\left<P_{\rm Puls}\right>$ & $\widetilde{P}_{\rm Puls}$ &
	     $\left<A_{\rm Puls}\right>$ & $\widetilde{A}_{\rm Puls}$  &
	     $\left<P_{\rm BL}\right>$ & $\widetilde{P}_{\rm BL}$ & 
	     $\left<A_{\rm BL}\right>$ &$\widetilde{A}_{\rm BL}$ & 
	     OoII \\
	   &  & (per cent) & (d) & (d) & (mag) & (mag) & (d) & (d) & (mag) & (mag) & (per cent) \\ 
		\hline
 All	&	2449	&	100	&	0.532	&	0.532	&	0.531	&	0.633	&	116	&	111	&	0.26	&	0.351	&	6.3\\
  a	&	2205	&	90	&	0.534	&	0.532	&	0.529	&	0.538	&	116	&	60.3	&	0.259	&	0.248	&	5.9\\
  b	&	95	&	3.9	&	0.492	&	0.49	&	0.545	&	0.541	&	107	&	69.44	&	0.399	&	0.4	&	0.0\\
  c	&	53	&	2.1	&	0.545	&	0.546	&	0.565	&	0.555	&	103	&	56.18	&	0.109	&	0.103	&	7.5\\
  d	&	41	&	1.7	&	0.517	&	0.52	&	0.563	&	0.559	&	93	&	65.66	&	0.165	&	0.161	&	0.0\\
  e	&	29	&	1.2	&	0.522	&	0.509	&	0.528	&	0.59	&	152	&	63.98	&	0.155	&	0.156	&	0.0\\
  f	&	26	&	1.1	&	0.506	&	0.499	&	0.523	&	0.519	&	170	&	134.34	&	0.4	&	0.377	&	0.0\\
		\hline
	\end{tabular}\label{Tab:MorphTypes}
\end{table*}

According to \citet{benko2011,Benko2018} and \citet{szeidl2012}, the variation of the light curve shape can be well approximated using notation of the data-transfer signal as the amplitude and phase/frequency modulation of the carrier wave. We aim to investigate purely amplitude modulation and the shape of the modulation envelopes not taking into account the period modulation. This would need the full frequency analysis of every particular star, which is clearly out of scope of this paper. Following \citet{Benko2018}, pure amplitude-modulated signal can be expressed as
\begin{equation}\label{Eq:Modulation}
    m^{\rm A}(t)=\left[ a^{\rm A}_{0} + f^{\rm A}_{\rm m}(\Omega, t)\right]m(t)
\end{equation}
where $f^{\rm A}_{\rm m}(\Omega,t)$ is the function that describes amplitude modulation with the angular frequency $\Omega=2\upi f_{\rm m}$, $a^{\rm A}_{0}$ and $a_{0}$ are the zero point constants and $m(t)$ is the non-modulated light curve (Eq.~\ref{Eq:FourSum}). The amplitude modulation can generally be complex and non-sinusoidal. In such case, $f^{\rm A}_{\rm m}(\Omega,t)$ can be expanded into a sine series and the shape of the envelope will depend on the amplitudes but also on the phases of the modulation components $\varphi_{i{\rm m}}^{A}$. 

\section{Results}\label{Sect:Results}

\subsection{Modulation periods}\label{Sect:ModulationPeriods}

\begin{figure}
	\includegraphics[width=\columnwidth]{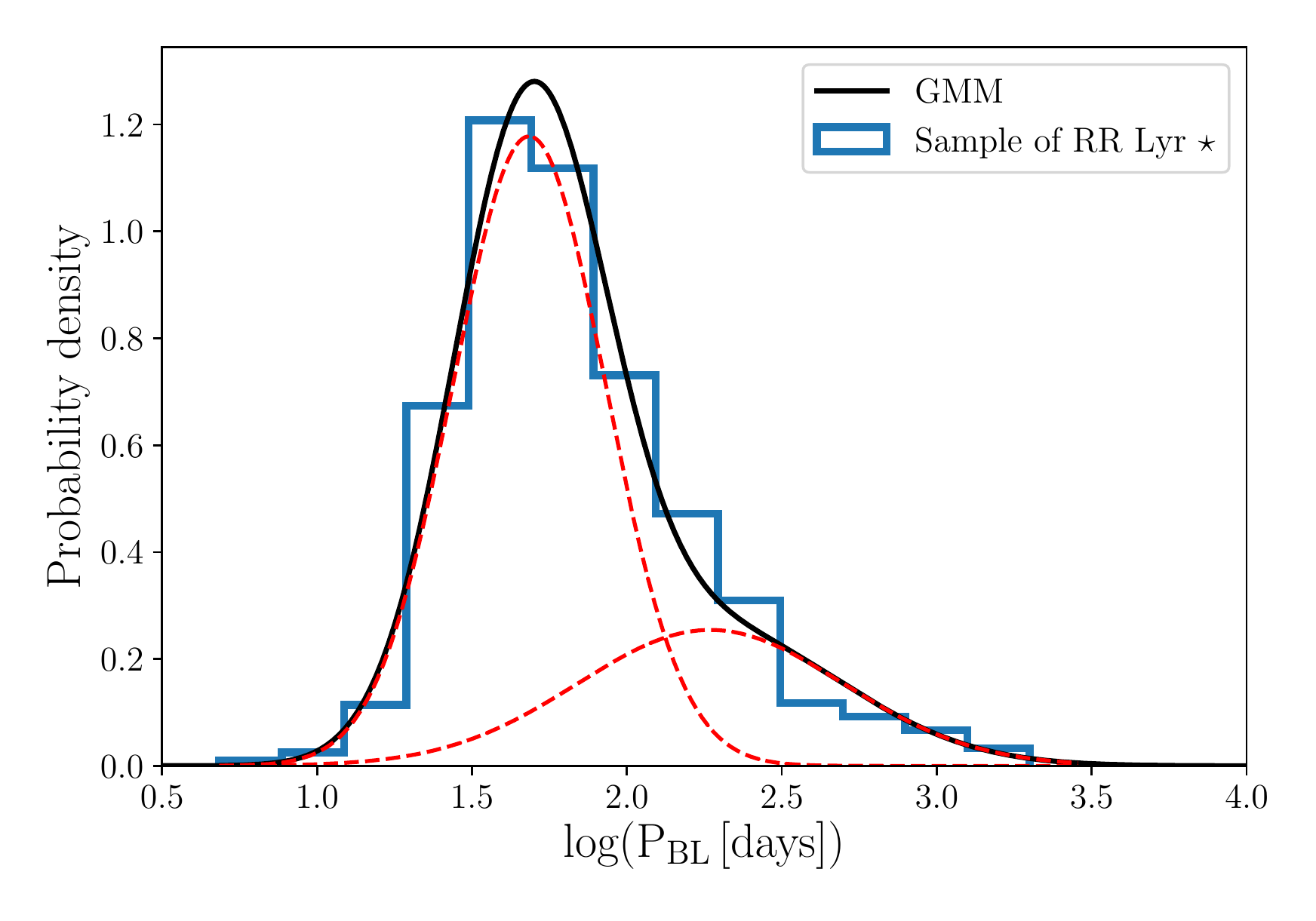}\\
	\includegraphics[width=.49\columnwidth]{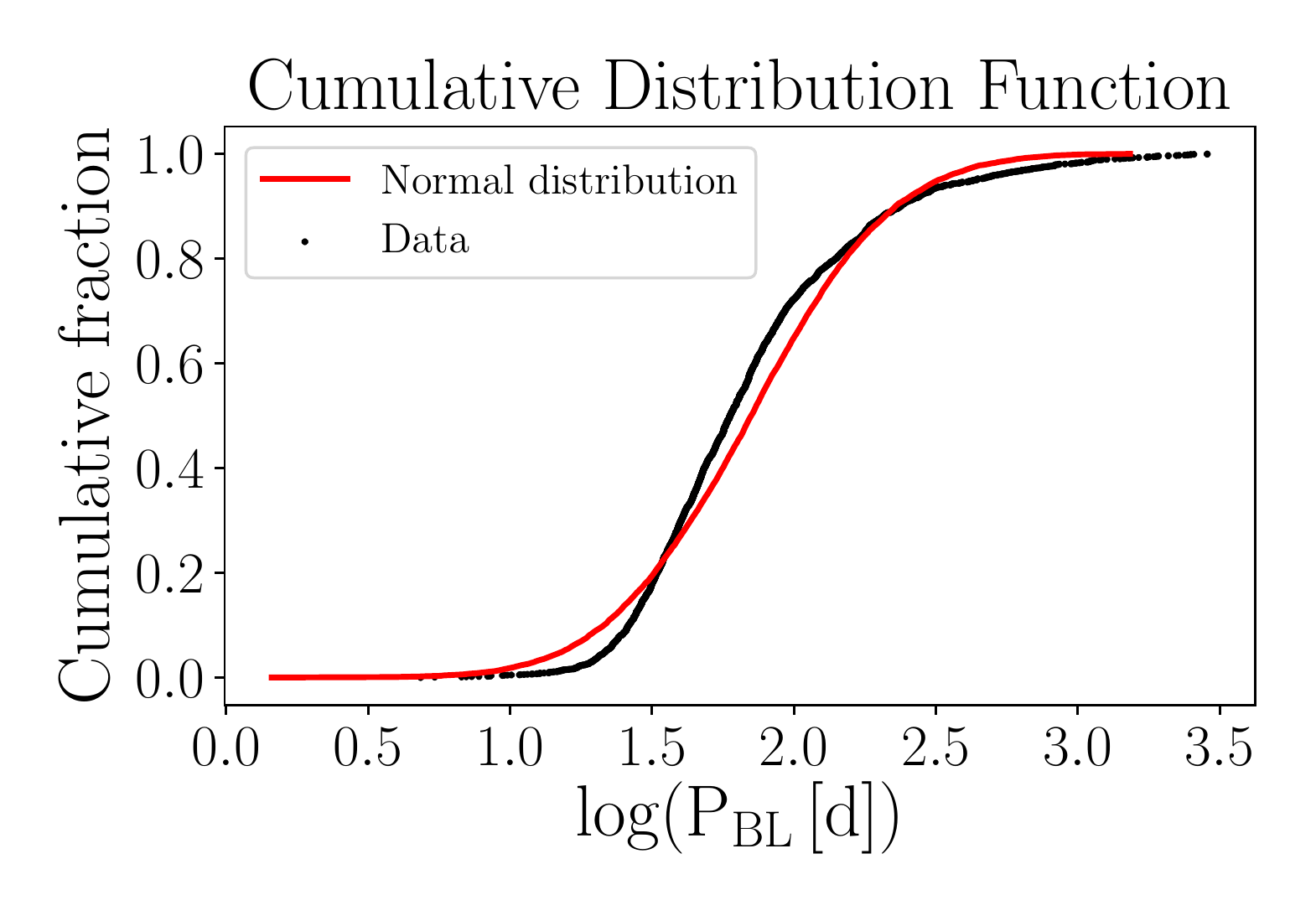}
	\includegraphics[width=.49\columnwidth]{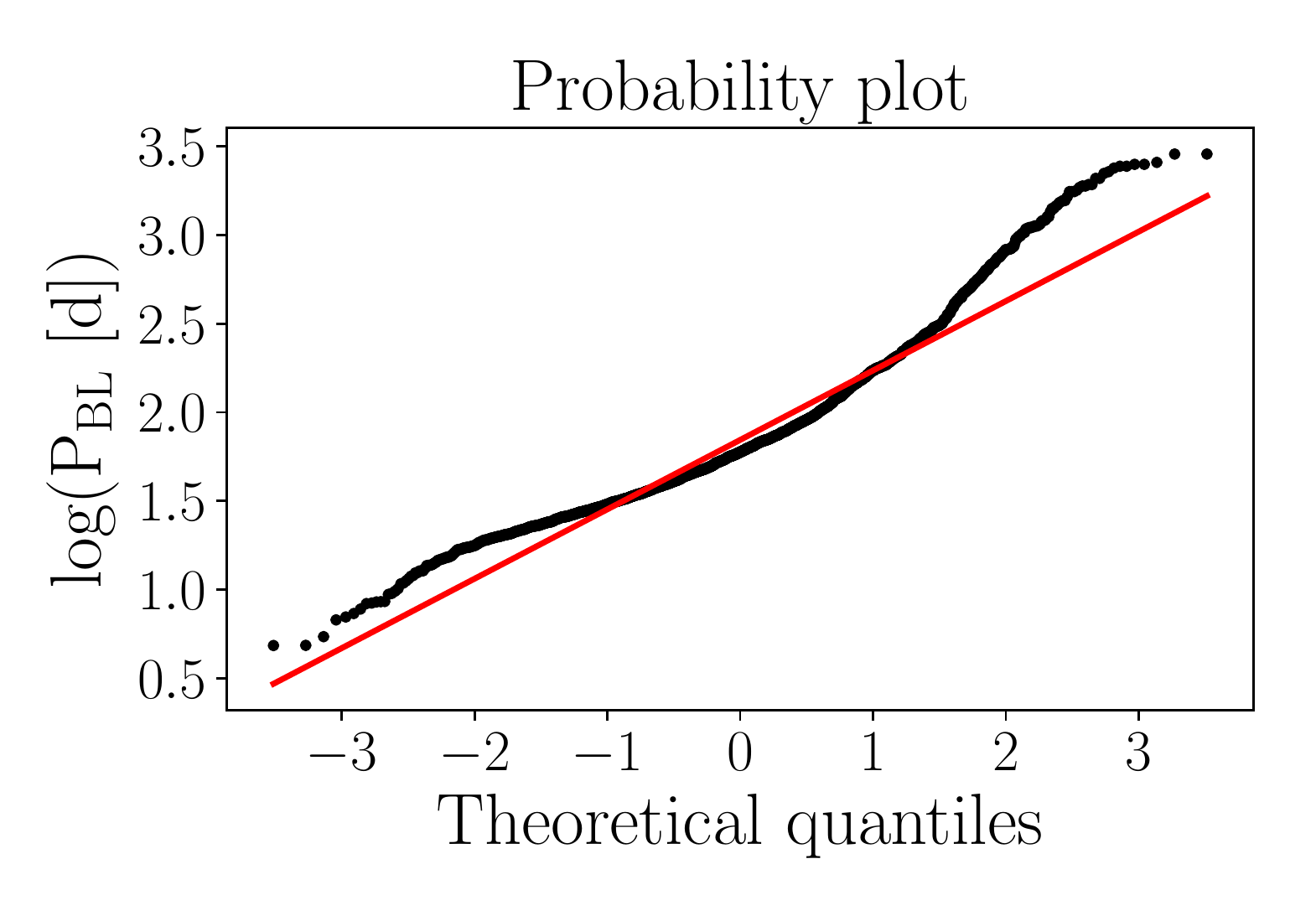}
    \caption{The distribution of the modulation periods of the sample stars modeled by the two Gaussian distributions (two dashed lines) and their sum (black solid line, {\it top panel}). The bottom panels show the cumulative distribution function ({\it bottom left}) and the probability plot ({\it bottom right}).}
    \label{Fig:BlazhkoPeriodDistribution}
\end{figure}

The distribution of the modulation periods is shown in Fig.~\ref{Fig:BlazhkoPeriodDistribution}. The shortest found \bl~period is 4.84\,days (OGLE-BLG-RRLYR-04885), which is the shortest known modulation period among fundamental-mode RRLs\footnote{The former record holder was SS Cnc with the modulation period of 5.3\,days \citep{jurcsik2006}. We note that OGLE-BLG-RRLYR-04885 has the side peak near the {\it SNR} limit for identifying it as Blazhko star.}. Among the first-overtone RRLs, the shortest ever detected modulation period is 2.23\,days \citep{netzel2018}. On the other hand, the longest detected modulation period among our sample stars is 2857\,days, which is far not the longest known value\footnote{\citet{jurcsik2016} found that V144 in the M3 globular cluster has a modulation period longer that 25 years.}. Median and average values of the modulation periods are 59.6\,days and 126\,days, respectively. The median value is basically the same as what was found by \citet{skarka2016b}. 

The distribution of the logarithm of the \bl~periods is not Gaussian as it is best apparent from the bottom panels of Fig.~\ref{Fig:BlazhkoPeriodDistribution}. We tested the non-normality of the distribution using statistical tests from the Python \texttt{Scipy} library \citep{Virtanen2020}. Both, the Kolmogorov--Smirnov and Shapiro--Wilk tests give the p-values orders of magnitude less than the critical value set to 1\,\% (0.01). The Anderson-Darling normality test also rejects the normality -- the statistical value 46.5 is much higher than the significance level for 1\,\% which is 1.091.

Although tests using the mean, median, standard error, skewness and kurtosis show that the distribution in Fig.~\ref{Fig:BlazhkoPeriodDistribution} is unimodal \citep{Rohatgi1989,Basu2006}, the cumulative distribution function and the probability plot shown in the bottom panels of Fig.~\ref{Fig:BlazhkoPeriodDistribution} suggest either multimodality or possibly also log-normal distribution. If we take the logarithm of the logarithm of the modulation period and perform the above mentioned normality tests, we come to the conclusion that such distribution is also unimodal and non-Gaussian but shows peculiarities. Thus, the log-normality cannot explain the features seen in the bottom panels of Fig.~\ref{Fig:BlazhkoPeriodDistribution}. 

In order to verify the existence of the two populations we used the \texttt{Gaussian mixture model} from the \texttt{scikit-learn} library \citep{Pedregosa2011}. First, we generated 1000 random realization of the Blazhko modulation periods assuming its error follows the Gaussian distribution. For each randomly generated distribution, we calculated the \texttt{Gaussian mixture model} with a varying number of Gaussians (from 1 to 10). Based on Bayesian information criterion (BIC) and Akaike information criterion (AIC), we estimated the suitable number of Gaussian components in each distribution. Both criteria suggested using at least 2-3 Gaussians to describe the distribution of modulation periods. We note that the errors in modulation periods (in Table~\ref{Tab:Parameters}) are estimated based on the Gaussian error propagation and are most likely underestimated. To reduce this effect in this analysis we artificially increased the errors by a factor of ten and run the \texttt{Gaussian mixture model} analysis again. 

Even with artificially increased errors, the two-population model with the mean modulation periods of 48\,days and 186\,days (shown by the different Gaussian fits represented by the red dashed lines in the top panel of Fig.~\ref{Fig:BlazhkoPeriodDistribution}) is the best result of the Gaussian mixture model analysis. The mean modulation periods are roughly in 1:4 ratio but this could be a coincidence.

\begin{figure}
	\includegraphics[width=\columnwidth]{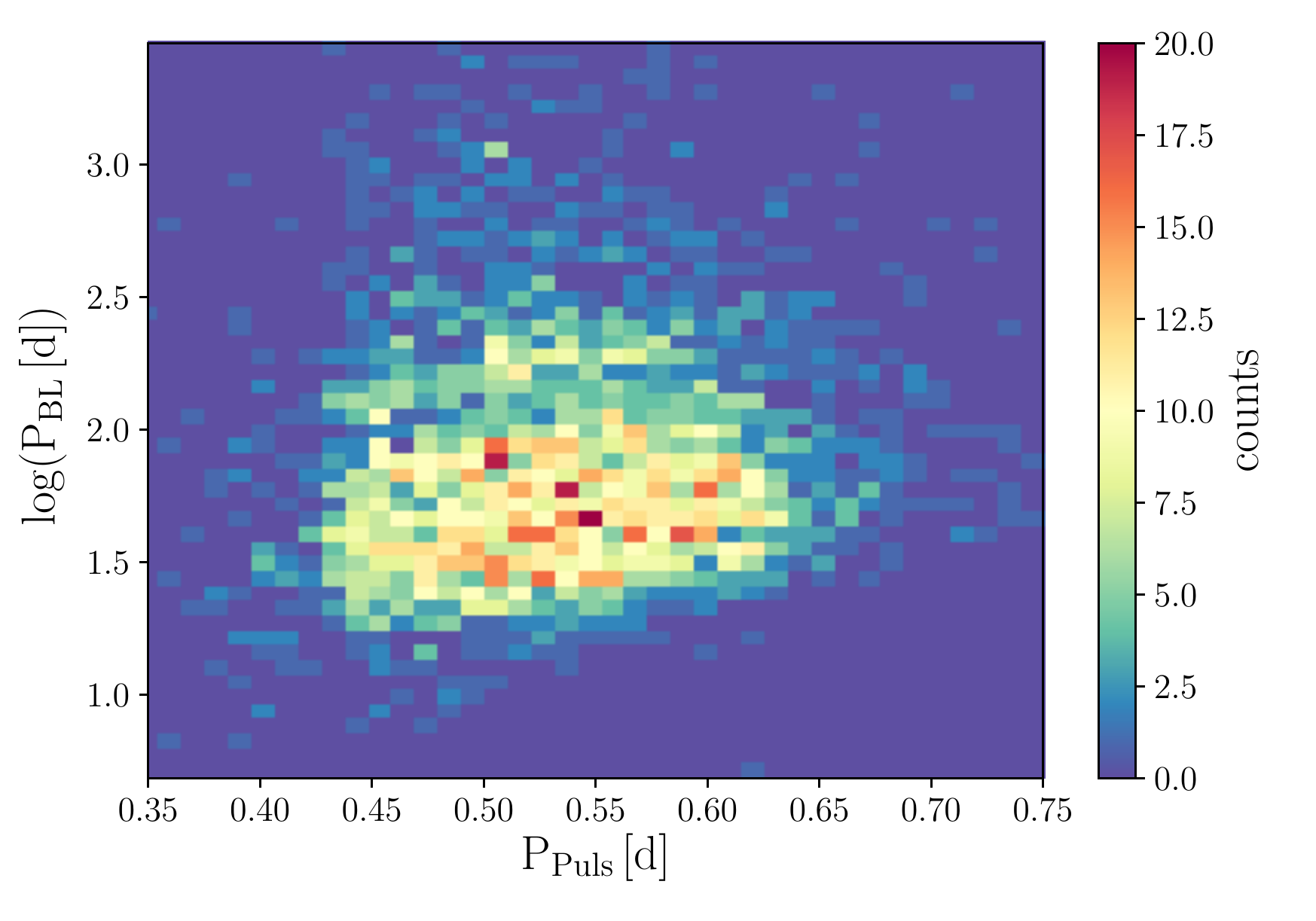}\\
	\includegraphics[width=\columnwidth]{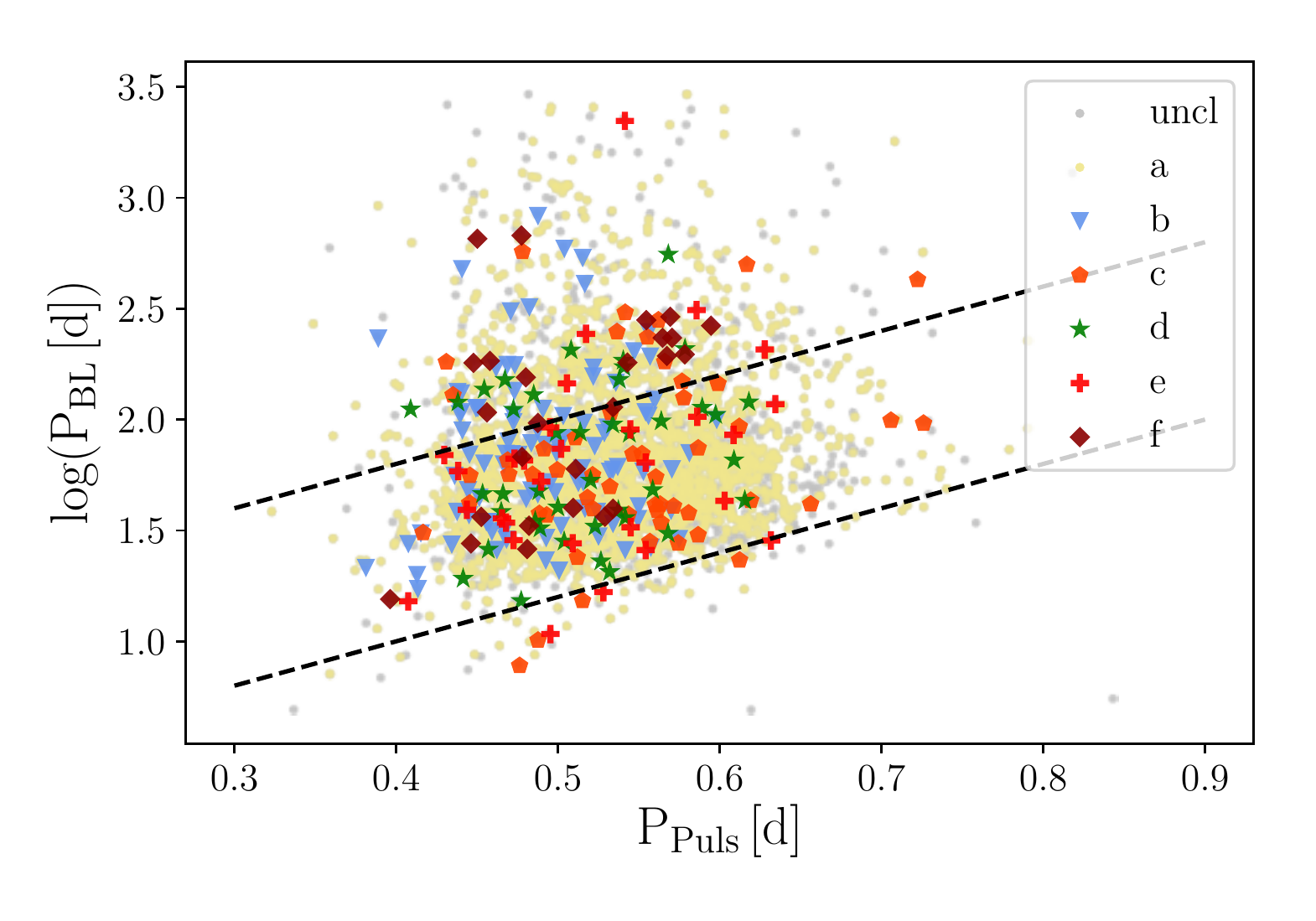}
    \caption{Density plot of the modulation periods ({\it top panel}). The bottom panel shows the modulation periods against pulsation period. Different modulation types are plotted with different symbols, the light-grey dots show unclassified stars. The bottom dashed line shows the limit of the modulation period with respect to the pulsation period ($\log$\pb$\approx2P_{\rm Puls}+0.2$), the upper dashed line shows the position of the Blazhko valley (the bottom line shifted +0.8 in $\log$\pb.}
    \label{Fig:DensityPlot}
\end{figure}

The two populations suggested from the Gaussian mixture model analysis are also apparent from the plots in Fig.~\ref{Fig:DensityPlot}. There is a valley separating the two populations in the density plot which we call {\it the Blazhko valley} (the top panel of Fig.~\ref{Fig:DensityPlot}). The center of the Blazhko valley is schematically shown by the upper dashed line in the bottom panel of the same figure. If the valley were horizontal, the two populations would be more distinct in the distribution of log$(P_{\rm BL})$.

The stars of the long-modulation period population (above the upper dashed line in the bottom panel of Fig.~\ref{Fig:DensityPlot}) have the spread of pulsation periods similar to the population below the Blazhko valley. Thus, the reason for their longer Blazhko periods can hardly account for the mean density, which mainly defines the pulsation period. It seems that the stars of all morphological types populate the whole distribution of the \pp vs. \pb, but the number of a-type stars decreases with longer modulation periods (the bottom panel of Fig.~\ref{Fig:DensityPlot}), while the distribution of other classes is more or less uniform. This results in more frequent occurrence of the non-sinusoidal modulation among stars with long modulation periods. We did not find any difference in the mean parameters between the long- and short-modulation period populations when the stars were separated based on the upper dashed line in the bottom panel of Fig.~\ref{Fig:DensityPlot}.

There are certain limits for the modulation periods. The bottom dashed line in the bottom panel of Fig.~\ref{Fig:DensityPlot} says that stars with short pulsation periods can have modulation periods of all lengths, while stars with long pulsation periods can have periods from a certain limit. For example, stars with \pp$>0.6$\,days can have \pb$>\sim20$\,days. This is exactly what \citet{jurcsik2005a} found. In the \pb versus \pp~plane, the liming line can be expressed as $\log$\pb$\approx2P_{\rm Puls}+0.2$. When this boundary line is shifted by 0.8 in log\pb, it falls exactly in the center of the Blazhko valley. The same slope of the line hints that there can really be some physical background for the valley.

\subsection{Modulation and mean brightness amplitudes}\label{Sect:ModulationAmplitudes}

\begin{figure}
	\includegraphics[width=\columnwidth]{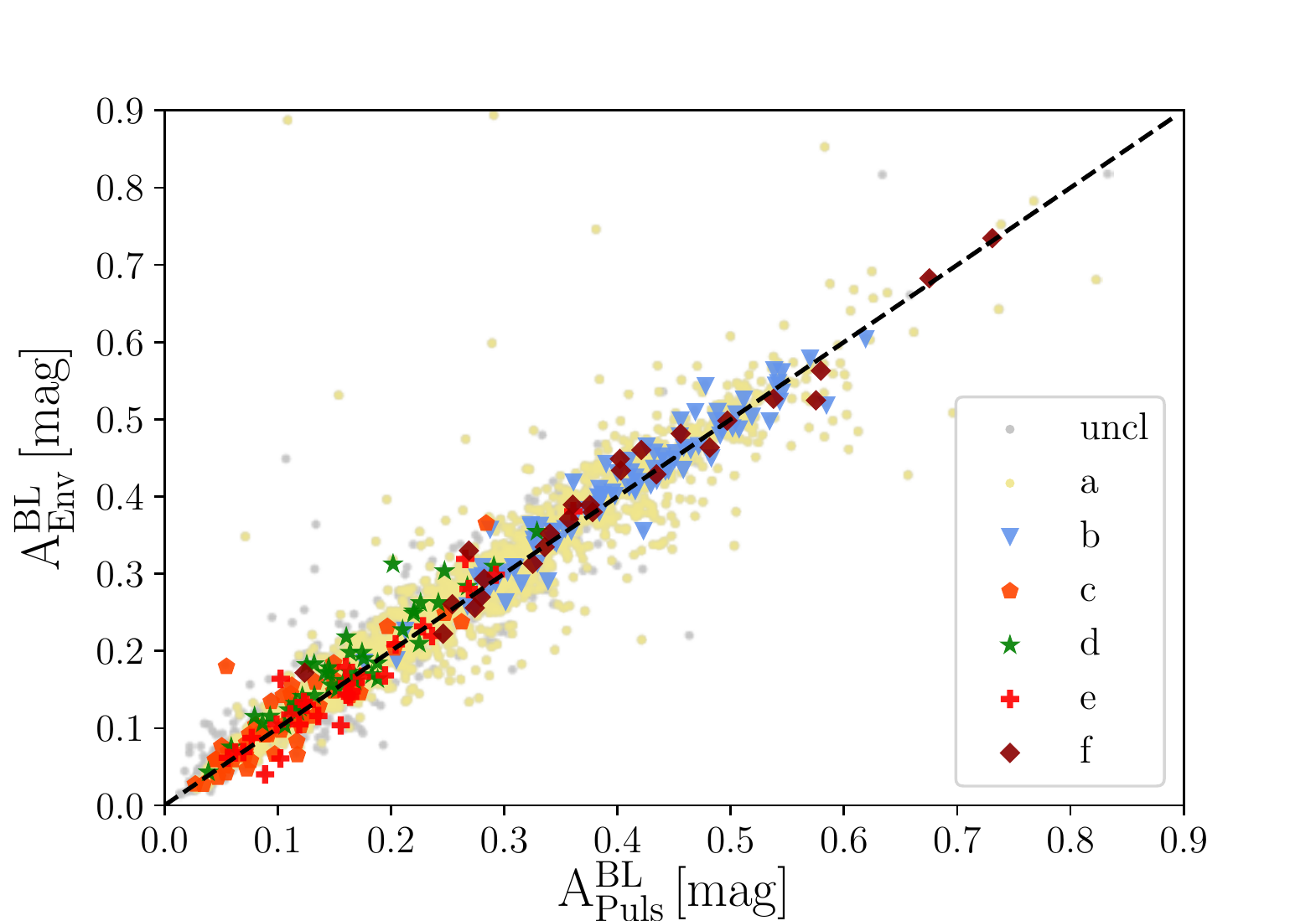}
    \caption{The comparison of the modulation amplitude determined on the basis of different approaches. The sum of amplitudes of the bottom and top envelopes is on the vertical axis, while the modulation amplitude based on the difference between total pulsation amplitude in Blazhko maximum and minimum phase is on the horizontal axis. The dashed black line shows the 1:1 correlation.}
    \label{Fig:ComparisonAmod}
\end{figure}

The modulation amplitude, which we determined using Eq.~\ref{Eq:AblPuls} and alternatively as a sum of the amplitudes of the top and bottom modulation envelopes (
Eq.~\ref{Eq:AblEnv}), corresponds to the strength of the amplitude modulation. The two approaches of determining the modulation amplitude are in great agreement and are, therefore, equally usable (Fig.~\ref{Fig:ComparisonAmod}). The determination of the modulation amplitudes failed in a few tens of stars because of the data sparseness and/or inadequate time distribution. The values for these stars are not reliable but are easily distinguishable as the outliers in Fig.~\ref{Fig:ComparisonAmod}.  

\begin{figure}
	\includegraphics[width=\columnwidth]{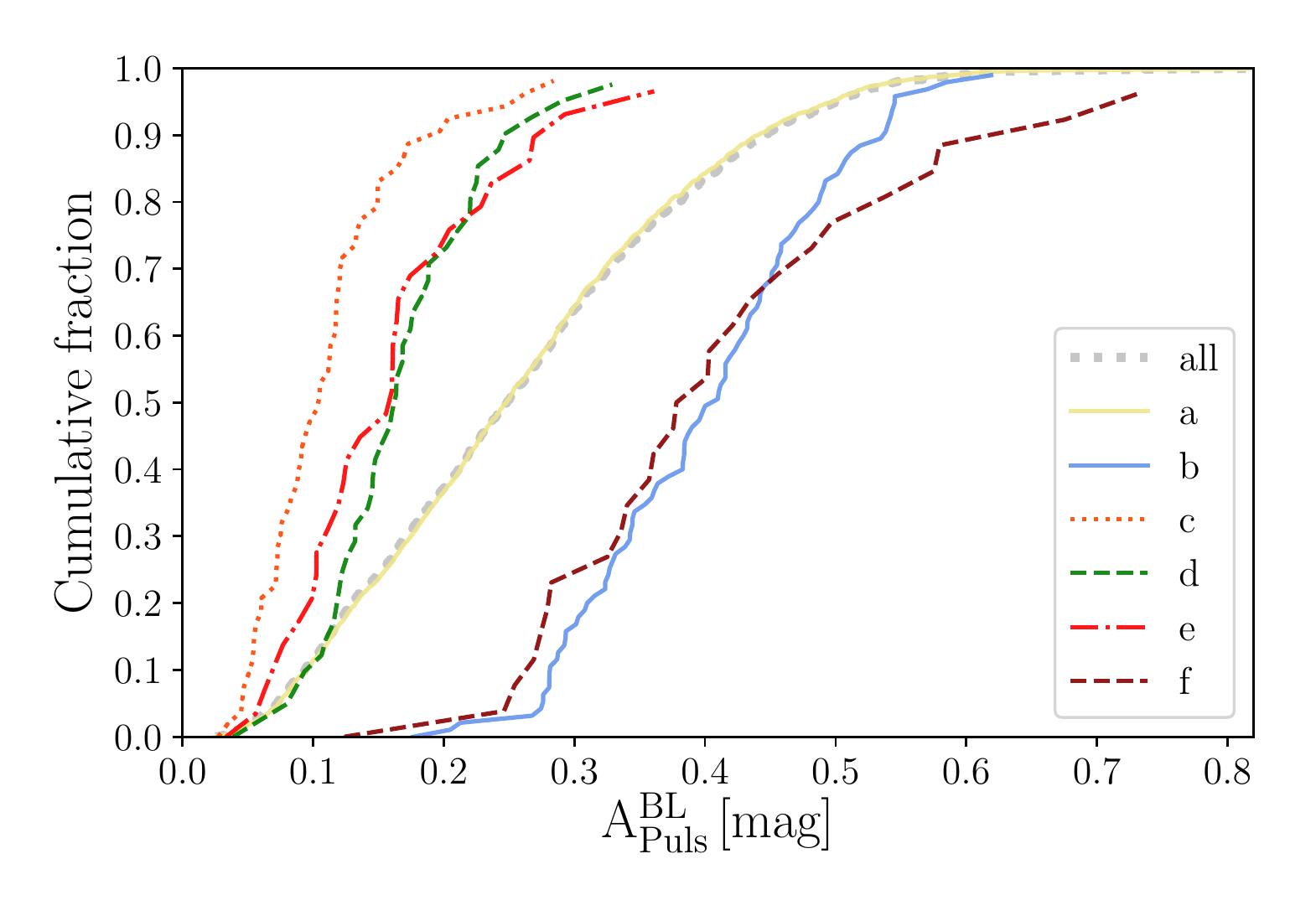}\\
	\includegraphics[width=\columnwidth]{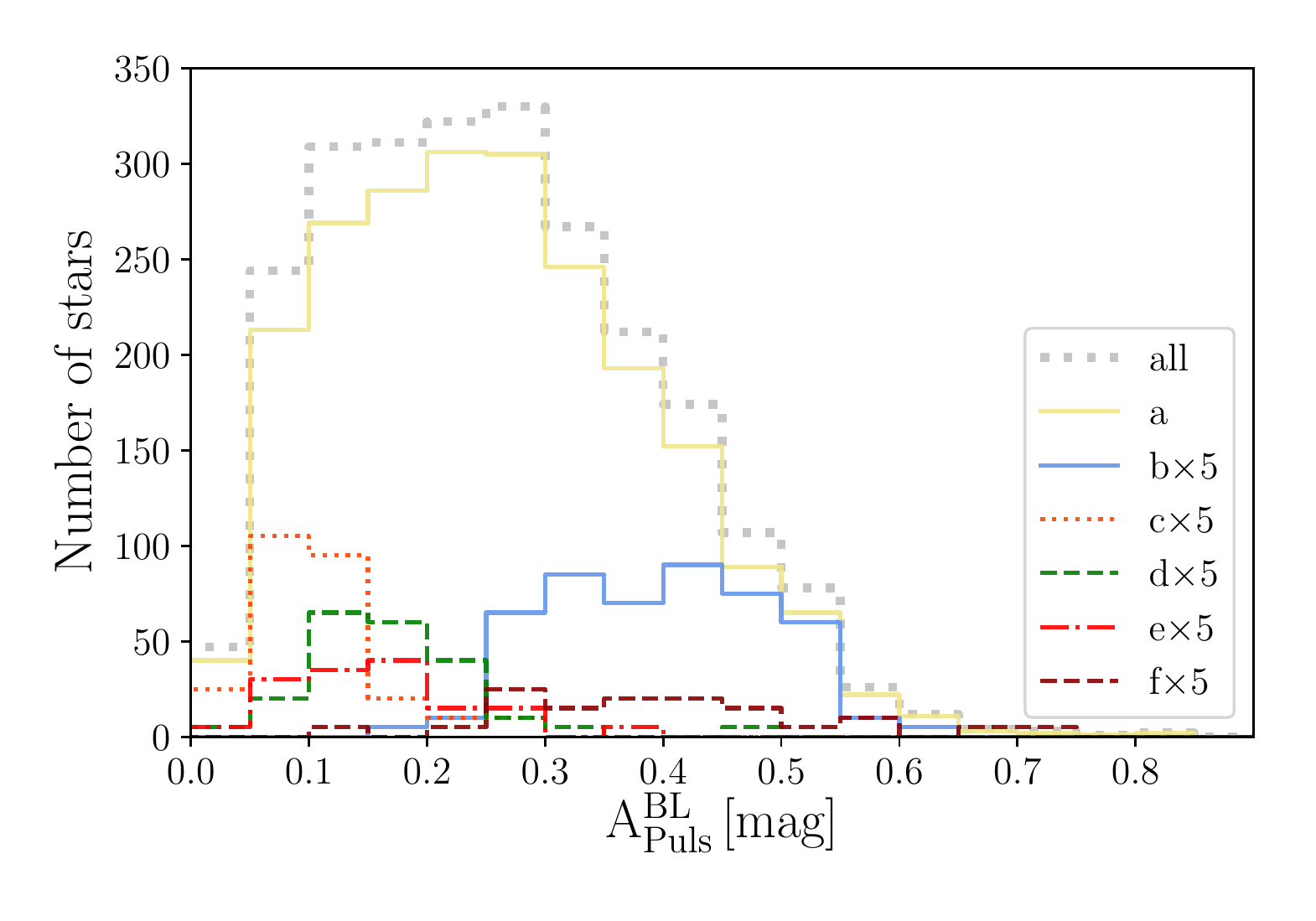}
    \caption{Cumulative distribution functions ({\it top panel}) and the distribution of the modulation amplitudes of the sample stars ({\it bottom panel}). The light-grey dotted curves show the full sample including non-classified stars.}
    \label{Fig:ModulationAmplitudes}
\end{figure}

The largest modulation amplitude is observed in OGLE-BLG-RRLYR-09437 ($A^{\rm BL}_{\rm Puls}=0.85$\,mag). This star shows two modulations which are in resonance. Thus, the maximal amplitude is very large, while the minimal amplitude is low. The smallest detected modulation amplitude is 0.0145\,mag in OGLE-BLG-RRLYR-05757. The cumulative distribution function (Fig.~\ref{Fig:ModulationAmplitudes}) shows that 99.2\,per cent of all the modulation amplitudes are below 0.6\,mag. Only very exceptionally, the amplitudes are higher. 

The overall distribution of amplitudes (Fig.~\ref{Fig:ModulationAmplitudes}) is defined mainly by stars from class a, which is the most populated (90~per cent of all stars). The stars from other types have very different modulation amplitudes which can be naturally expected because of the definition of the classes (Sect.~\ref{Sect:MorphologicalTypes}). Stars of b and f types have the largest amplitudes, while type-c stars have the smallest amplitudes (see both panels of Fig.~\ref{Fig:ModulationAmplitudes} and Fig.~\ref{Fig:AblVsPbl}). Note that types b and f and types d and e have very similar distributions. 

\begin{figure}
	\includegraphics[width=\columnwidth]{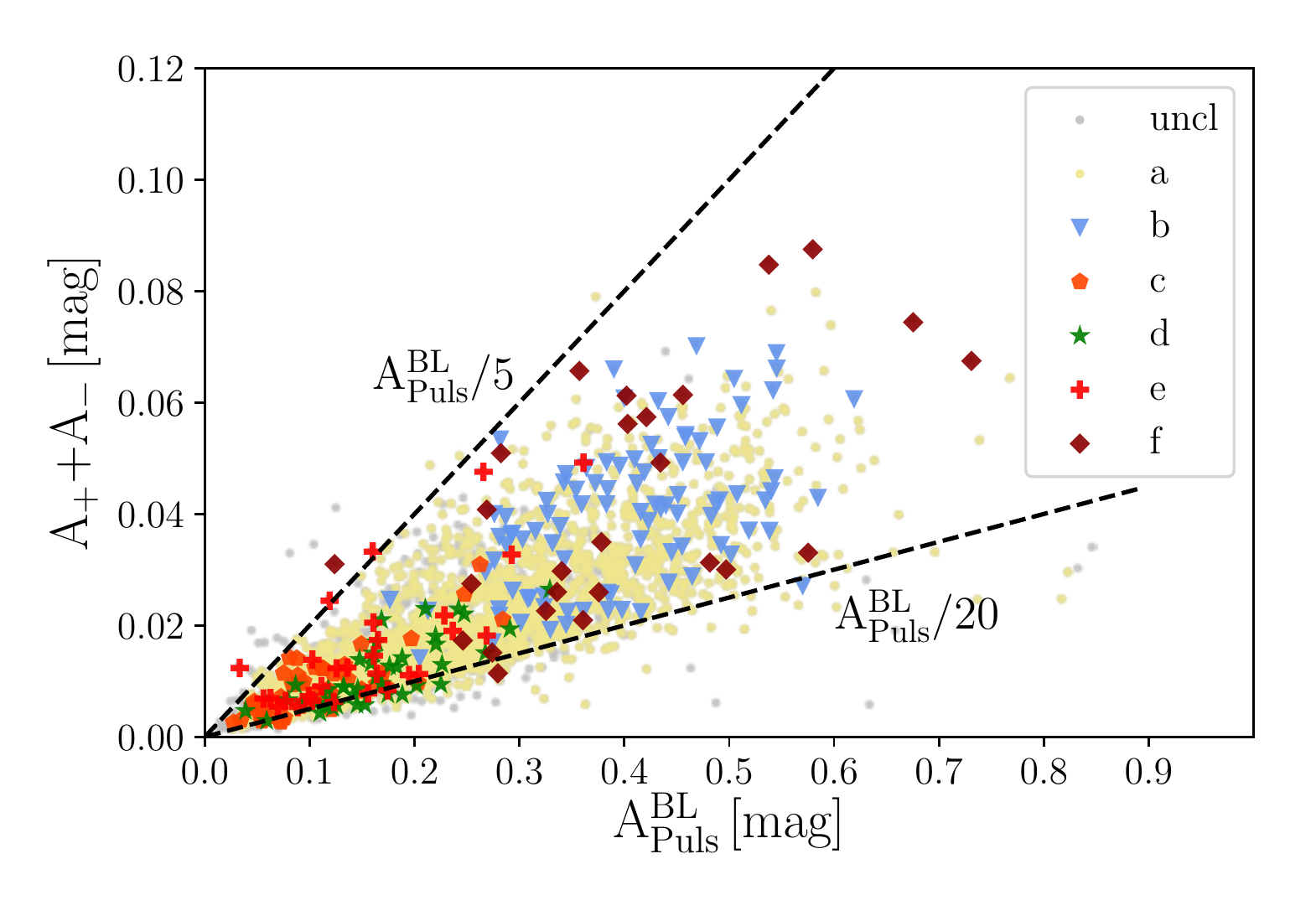}
    \caption{Dependence of the sum of the side peak amplitudes vs modulation amplitude. The dashed lines show the total Blazhko amplitude $
    A^{\rm BL}_{\rm Puls}$ divided by 5 and 20, respectively.}
    \label{Fig:Apeaks_Abl}
\end{figure}

Definition of the modulation amplitude based on the amplitudes of the frequency side peaks, as used and discussed by \citet{jurcsik2005a} and \citet{Benko2014b}, can be difficult because the amplitudes of the side peaks (or their sum) are not exactly proportional, nor equal to the modulation amplitude. The quality and distribution of the data, as well as the presence of the additional amplitude and/or phase modulation influence the amplitudes of the frequency peaks \citep{jurcsik2005c,benko2011}. We demonstrate this issue in Fig.~\ref{Fig:Apeaks_Abl}. The figure shows how the sum of the amplitudes of the $f_{0}\pm f_{m}$ peaks depends on the modulation amplitude determined from Eq.~\ref{Eq:AblPuls}. The sum of the amplitudes of the $f_{0}\pm f_{m}$ peaks is 5 to 20 times smaller than the modulation amplitude.

\begin{figure}
	\includegraphics[width=.5\columnwidth]{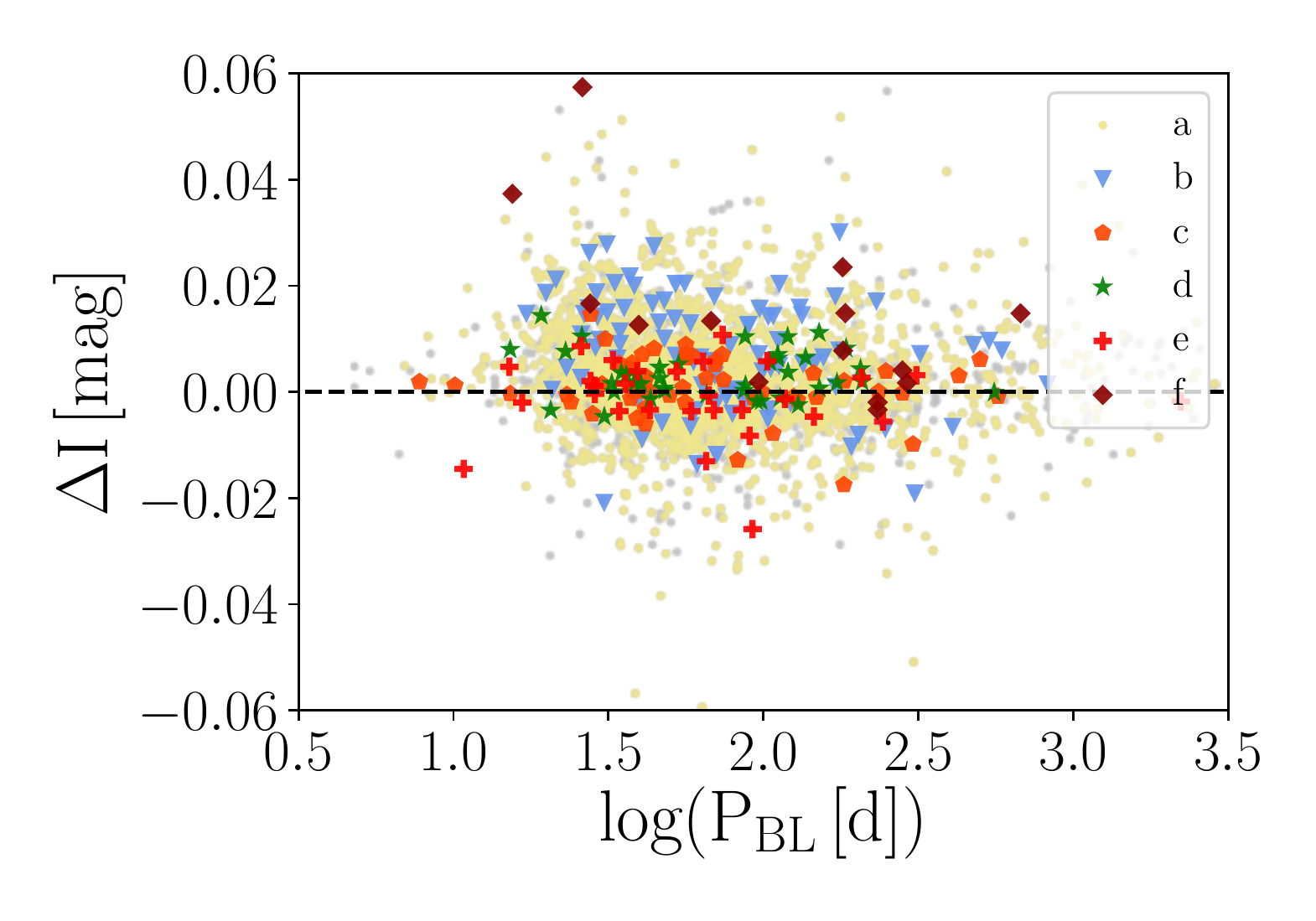}\includegraphics[width=.5\columnwidth]{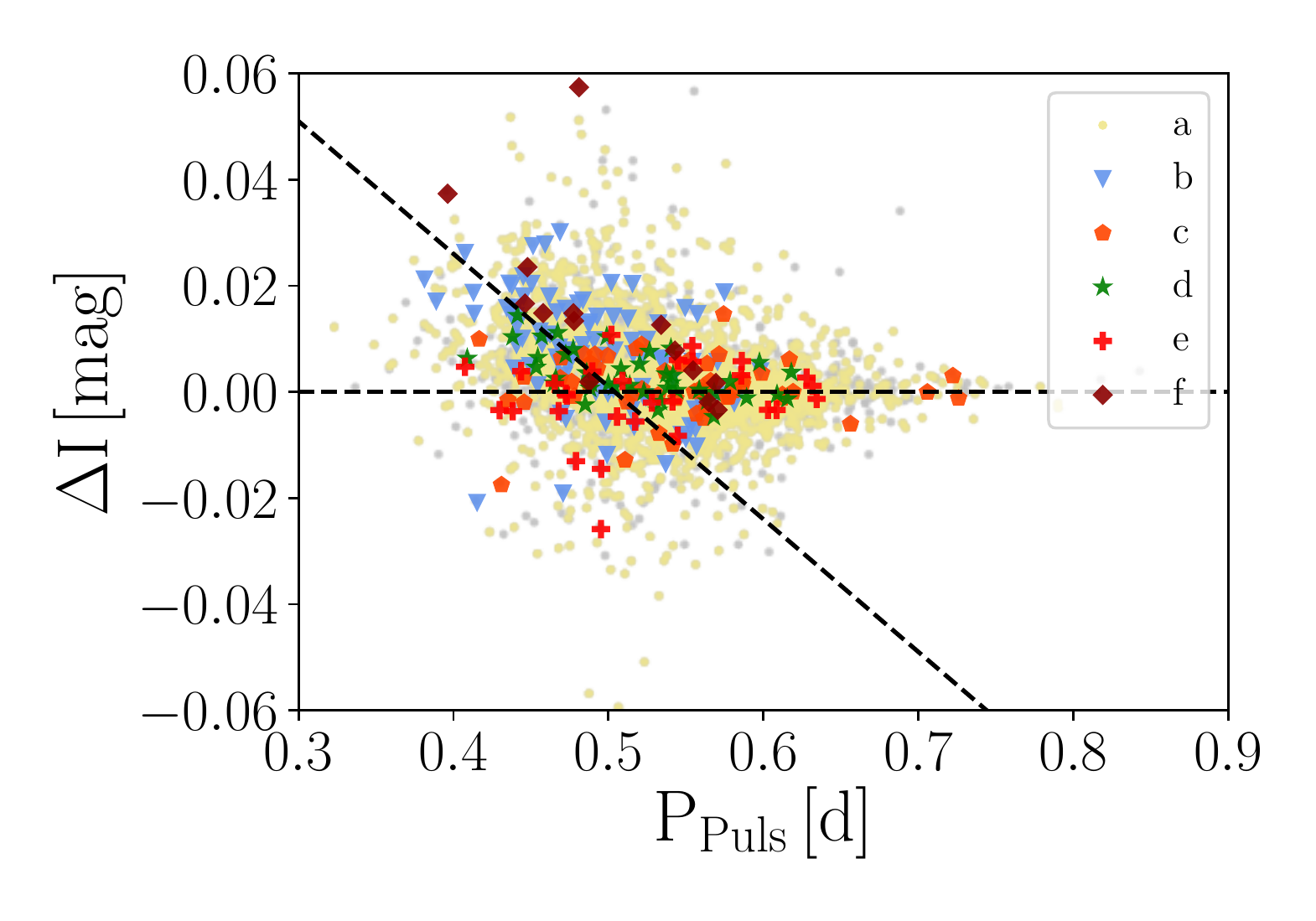}\\
	\includegraphics[width=.5\columnwidth]{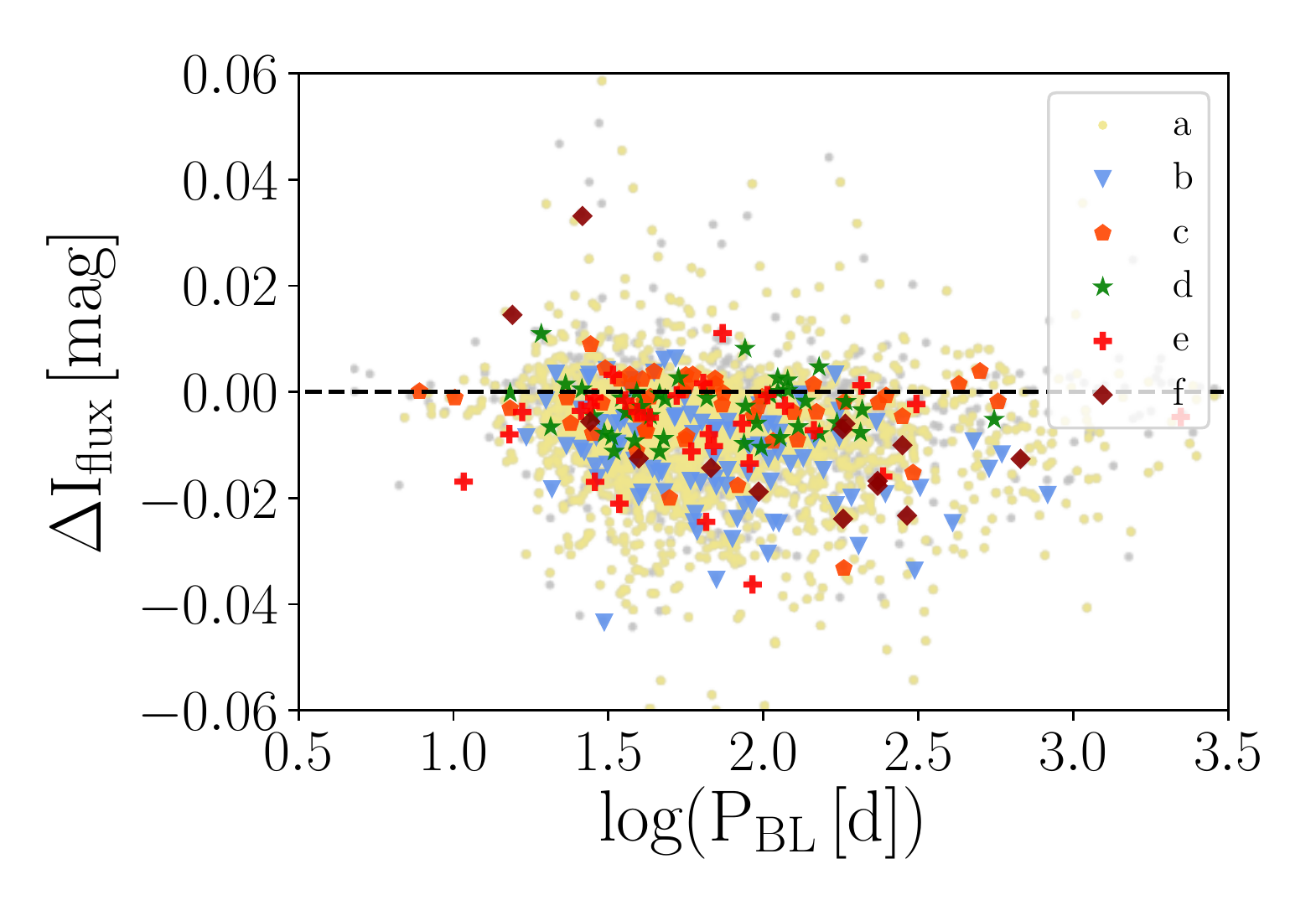}\includegraphics[width=.5\columnwidth]{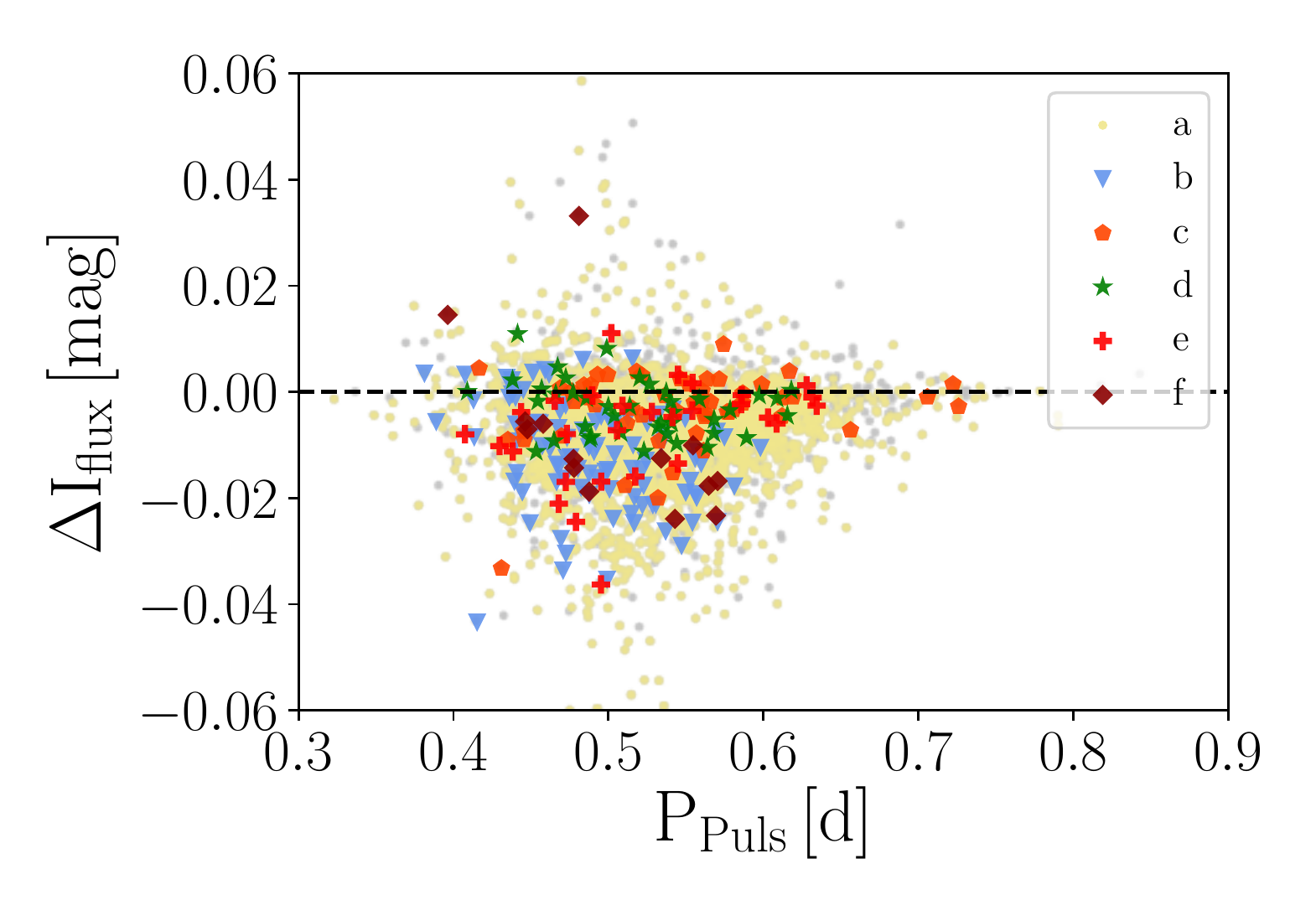}\\
    \caption{Dependence of the amplitude of the mean magnitude variation $\Delta I$ versus modulation ({\it left panels}) and pulsation period ({\it righ-hand panels}). The negative values mean that the mean brightness during minimal Blazhko phase is larger than during maximal Blazhko phase (see the definition of $\Delta I$ in Eq.~\ref{Eq:MeanVariation}). The top panels show the $\Delta I$ variation based simply on the OGLE-IV magnitudes, the bottom panels show $\Delta I_{\rm flux}$ when the magnitudes are transformed to fluxes, the difference is calculated and transformed back to magnitudes.}
    \label{Fig:MeanVariation}
\end{figure}

\begin{figure}
	\includegraphics[width=\columnwidth]{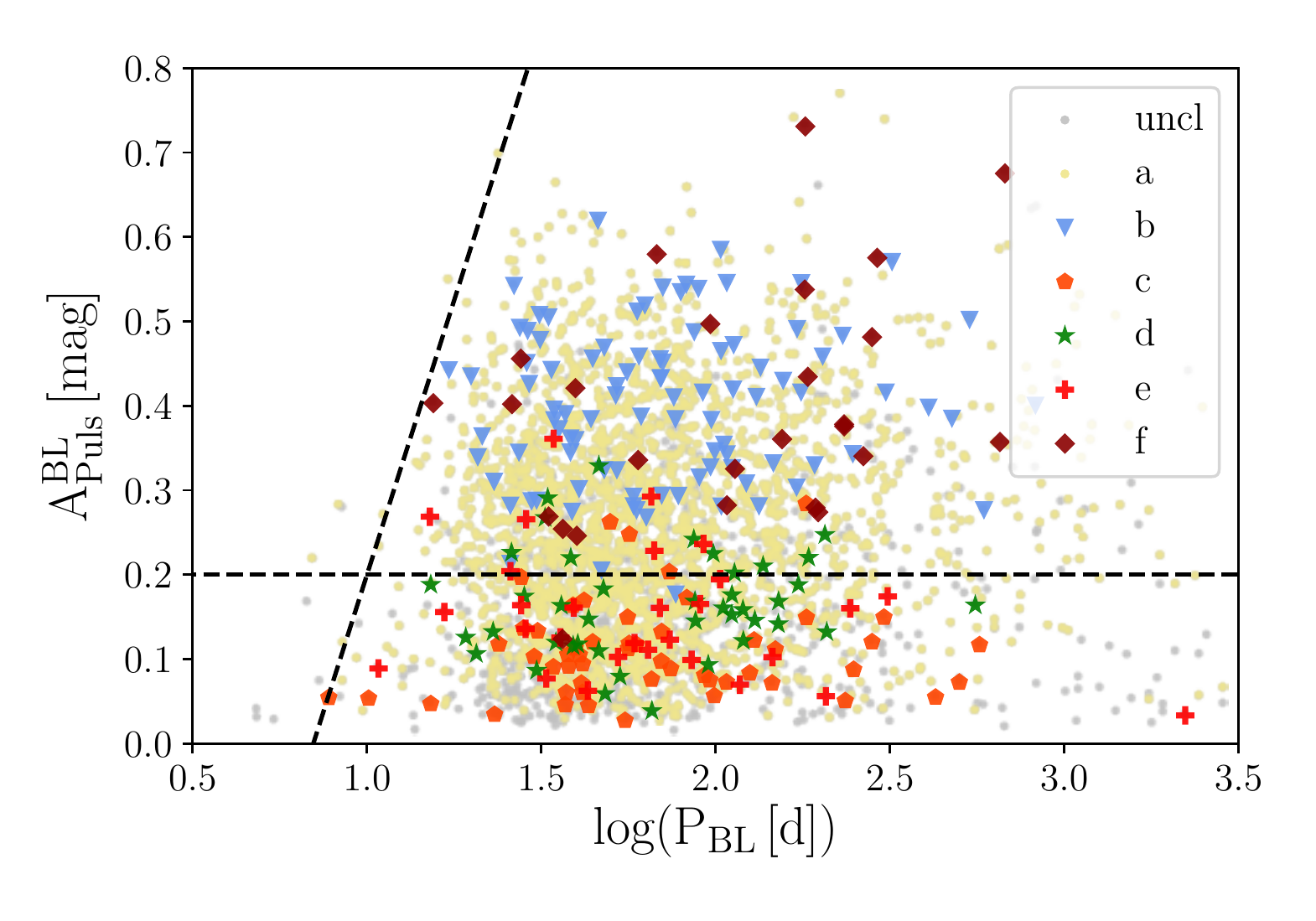}
    \caption{Dependence of the amplitude of the modulation versus modulation period. The tilted black dashed line shows the upper limit for the amplitude in the range of modulation periods of 0-30\,days. The slope is 1.3. The horizontal line shows the 0.2-mag limit separating b- and f-type stars from the others.}
    \label{Fig:AblVsPbl}
\end{figure}

Variation of the mean magnitude produces peak with frequency equivalent to the modulation frequency in the mmag regime. \citet{benko2014a} found in the {\it Kepler} data that the amplitude of this peak is proportional to the modulation period. We can test it by plotting the amplitude of the mean-magnitude variation $\Delta I$ (Eq.~\ref{Eq:MeanVariation}). We did not find any dependence even among stars with the largest number of points with the most reliable determination of $\Delta I$ (Fig.~\ref{Fig:MeanVariation}). 

Because the impact of the modulation is the most apparent in the variation of the upper modulation envelope in most of the stars, it can be naturally expected that most of the stars will be brighter during the maximum Blazhko amplitude ($\Delta I<0$). The opposite is true (the top panels of Fig.~\ref{Fig:MeanVariation}). Thus, test with fluxes was performed. First we transformed magnitudes into fluxes, calculated the difference in mean fluxes and transformed the difference back to magnitudes. The result is shown in the bottom panels of Fig.~\ref{Fig:MeanVariation}. Now the situation is as can be logically expected. This test reveals that using magnitudes can be misleading and can can give strange results. 
The difference in $\Delta I$ looks completely different when it is determined on the basis of OGLE magnitudes and when it is calculated from the flux-transformed values.

The study by \citet{Benko2014b} suggests that the amplitude of the modulation envelope is monotonically proportional to the modulation period. In other words, the longer the Blazhko period, the larger the amplitude of the modulation envelope\footnote{Similarly as for the amplitude of the low-frequency peak raised by the variation of the mean brightness.}. They note that large data sample is desired for confirmation of what they found, which is exactly our case. In Fig.~\ref{Fig:AblVsPbl}, we show that there is no monotonic dependence between the modulation amplitude and the modulation period. We can only find a hint of some limit (the tilted dashed line in Fig.~\ref{Fig:AblVsPbl}) saying that stars with Blazhko periods shorter than about 30\,days ($\sim1.5$ in logarithmic scale) can only have certain modulation amplitudes. It is interesting to point out that b- and f-type stars have predominantly amplitudes larger than 0.2\,mag, while those of c-e types below 0.2\,mag.

\begin{figure}
	\includegraphics[width=\columnwidth]{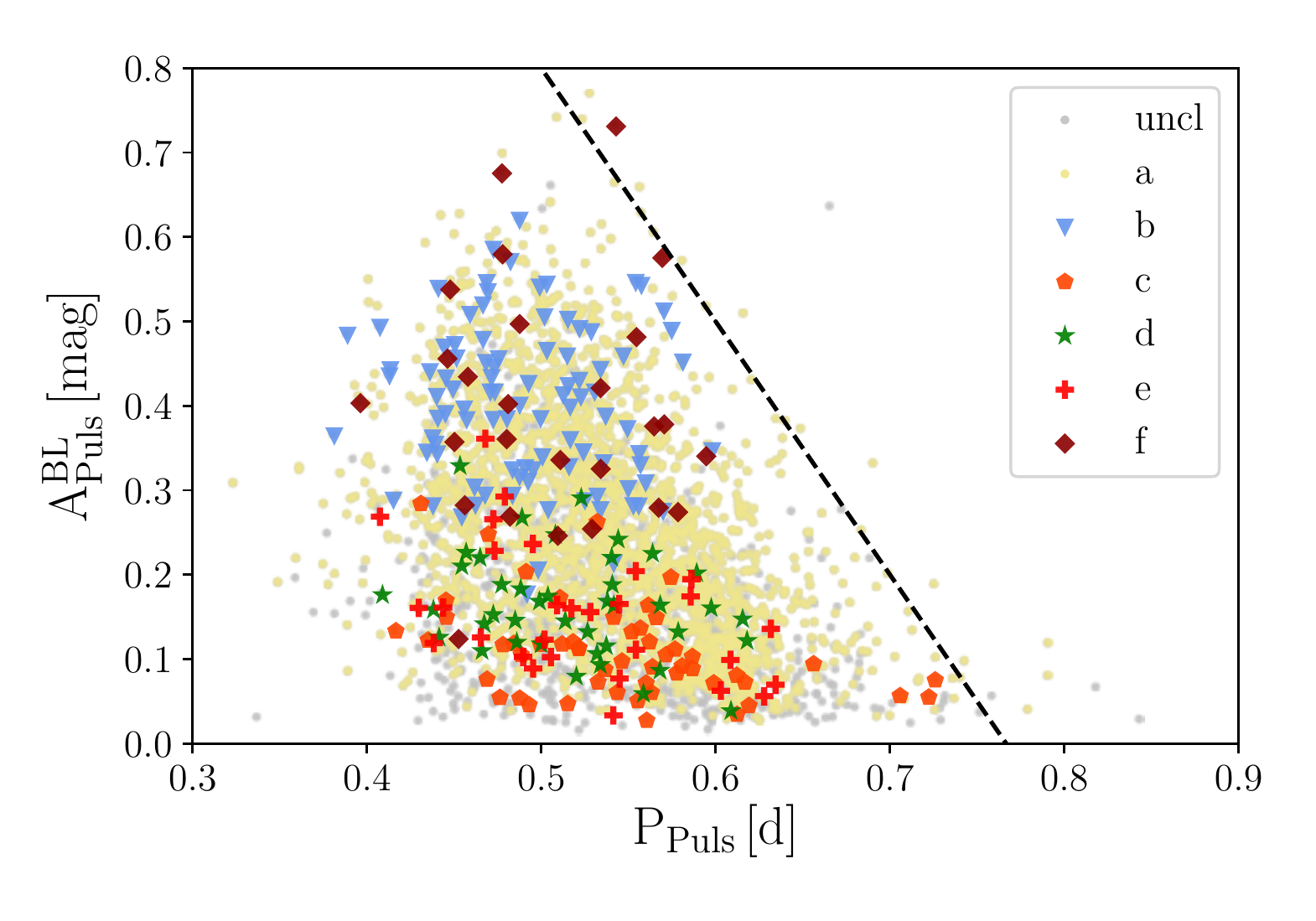}
    \caption{Dependence of the amplitude of the modulation versus pulsation period. The black dashed line shows the upper limit for the amplitude in the range of modulation periods of 0.5 to 0.9\,days. The slope is -3.}
    \label{Fig:AblVsPpuls}
\end{figure}

Figure~\ref{Fig:AblVsPpuls} shows that there is also top limit for modulation amplitudes for stars with pulsation periods above 0.5\,days (shown schematically with the dashed line). The modulation amplitude clearly decreases with increasing pulsation period, which was already observed in stars from globular clusters, galactic field and LMC by \citet{jurcsik2005a}. This hints towards that in RRLs with long pulsation periods, which are generally cooler, larger and more luminous than the short-period RRLs, some effect exist that suppress the amplitude of the Blazhko effect. 

Because RRLs with large modulation amplitudes have short pulsation periods, we can naturally expect that the stars with short pulsation periods will have larger variation in the mean brightness, which is really observed (see the bottom panel of Fig.~\ref{Fig:MeanVariation}). 

\subsection{Frequency spectra and multiple modulation}\label{Sect:FrequencySpectra}

We investigated only the vicinity of the basic pulsation frequency $f_{0}$. The difference between the $f_{+}$ and $f_{-}$ components (Eq.~\ref{Eq:df}) is for 99.7\,per cent of the stars ($3\sigma$) smaller than 0.00016\,c/d. The distribution of the asymmetry parameter $Q$ (Eq.~\ref{Eq:Q}) shows that 70~per cent of the stars has larger peak at the right-hand side of $f_{0}$ ($f_{+}$) than at the left side $f_{-}$. All morphological classes have similar statistics, but type d, where only 14.6~per cent of the stars show larger peak at higher frequency. Probably it could reflect the double minimum envelope of the modulation.

\begin{figure}
	\includegraphics[width=\columnwidth]{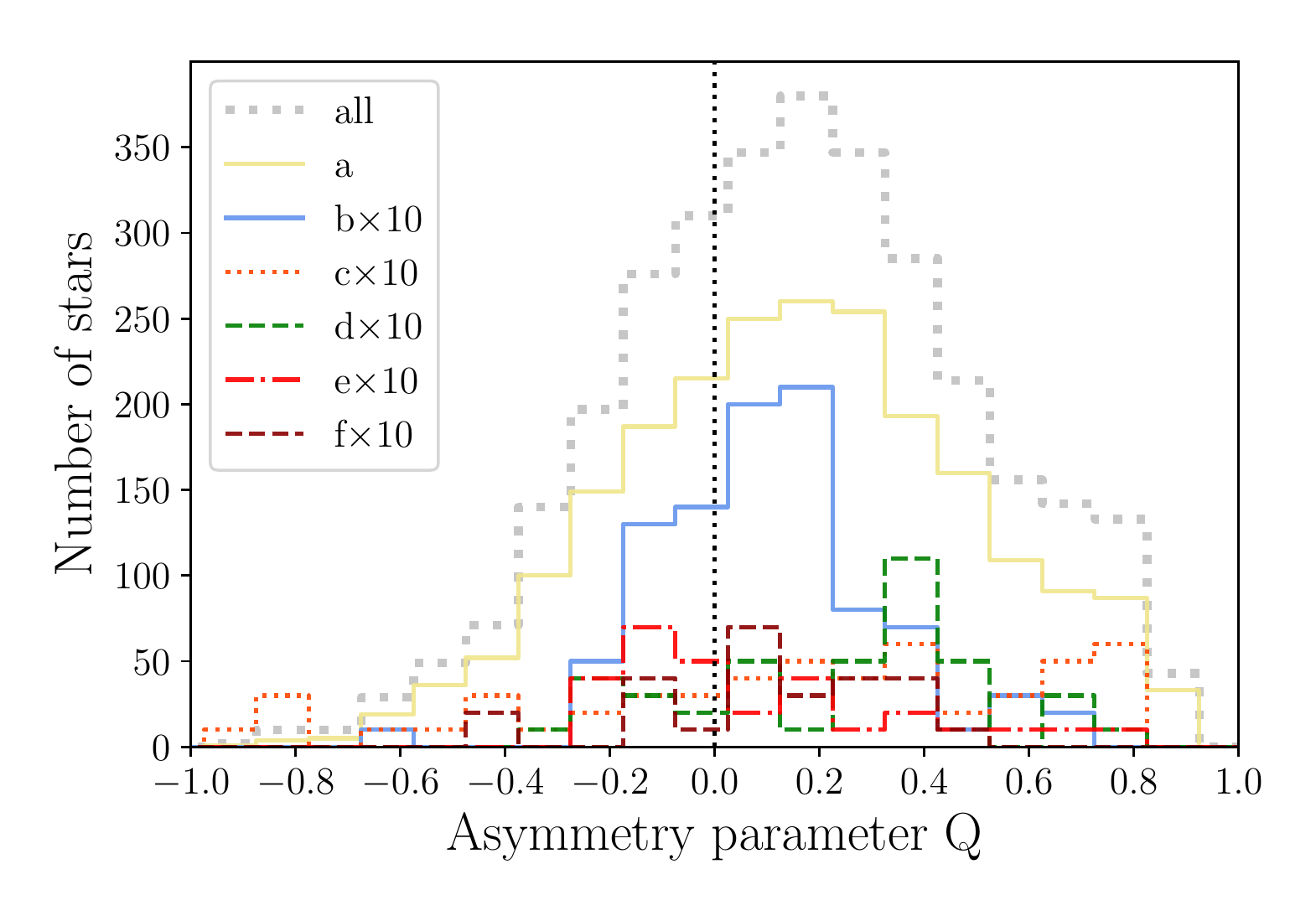}
    \caption{The distribution of the  asymmetry parameter $Q$. Almost 70\,per cent have $Q>0$ (the vertical dashed line).}
    \label{Fig:Q}
\end{figure}

According to the mathematical description of the modulation presented by \citet{benko2011}, peak with larger amplitude at the higher frequency side means that the phase difference $\phi_{\rm m}$ between amplitude and phase modulation is $\upi<\phi_{\rm m}<2\upi$. Such stars show counter-clockwise movement in the maximum brightness versus maximum phase diagrams. Our analysis shows that 70\,per cent of all Blazhko stars in GB comply this. \citet{alcock2003} found similar value (74\,per cent) based on the study of 731 stars in the Large Magellanic Cloud.  

About 25\,per cent of the studied stars (799 stars marked with AdP in Table~\ref{Tab:Parameters}) show additional modulation which we deduce from the additional peaks present in the vicinity of $f_{0}$. All morphological classes show similar percentages of stars with additional peaks, but type b. Stars with additional peaks in this group constitute only 11.6\,per cent of all the b-type stars. Some stars (at least 25) show also unresolved peaks that could mean period instability. 

\subsection{Asymmetry of the envelopes}

\begin{figure}
	\includegraphics[width=\columnwidth]{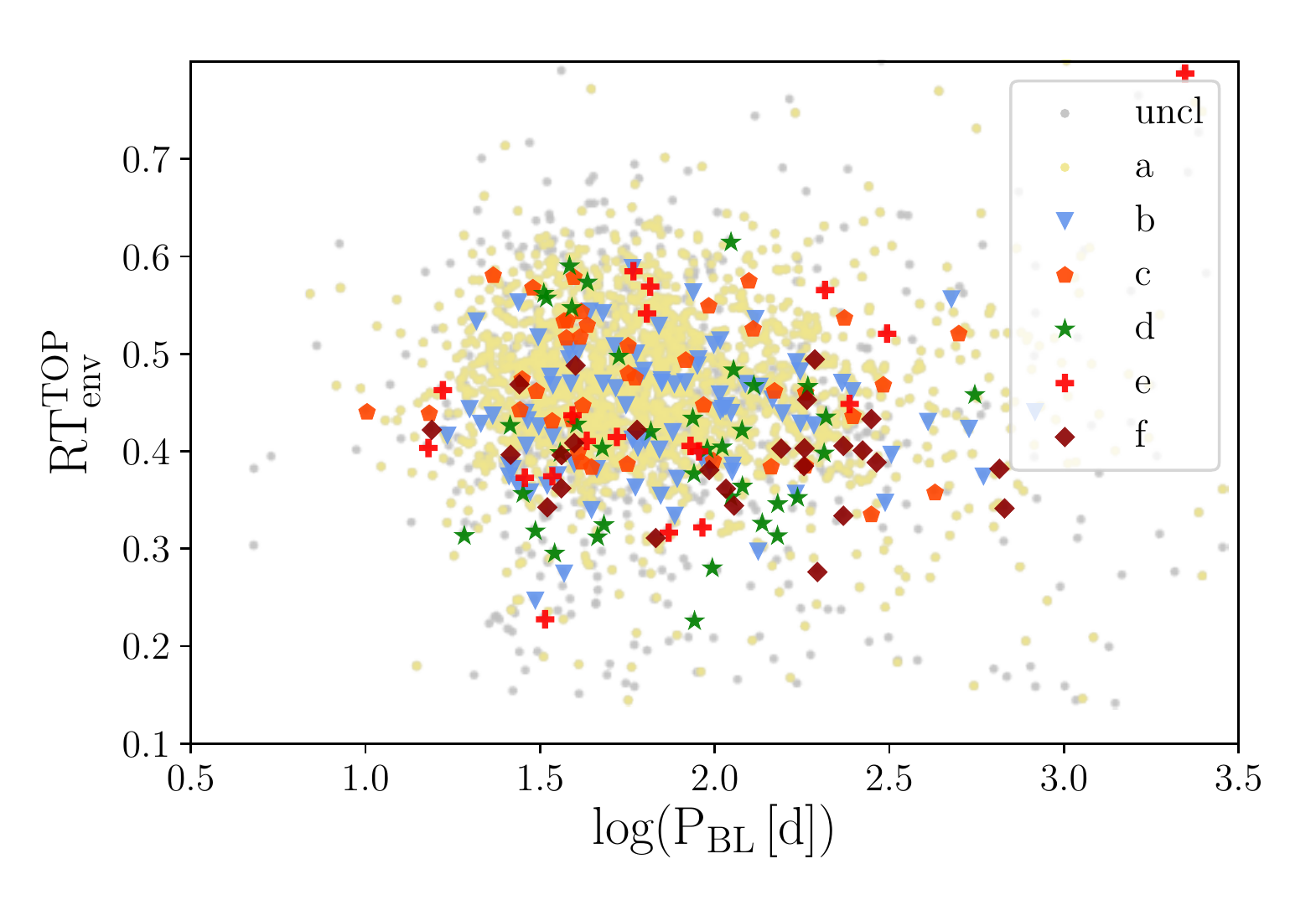}
    \caption{The rise time of the top modulation envelope. Sixty per cent of stars have {\it RT$^{\rm TOP}_{\rm env}$}$<0.5$.}
    \label{Fig:RTtopVsPbl}
\end{figure}

The asymmetry of the modulation envelopes can be deduced from the {\it RT$^{\rm TOP}_{\rm env}$} parameter of the envelope (Eq.~\ref{Eq:RT}, Fig.~\ref{Fig:RTtopVsPbl}). For 60\,per cent of all studied stars it is less than 0.5 which means that 60\,per cent of the Blazhko stars has asymmetric envelopes with steep rise to maximum. In case of class-b stars it is 87\,per cent and all the members of f-class show the steeper envelope rise to the maximum. 

\subsection{Fourier coefficients of the pulsation light curve}\label{Sect:FourCoefficients}

\begin{figure}
	\includegraphics[width=.5\columnwidth]{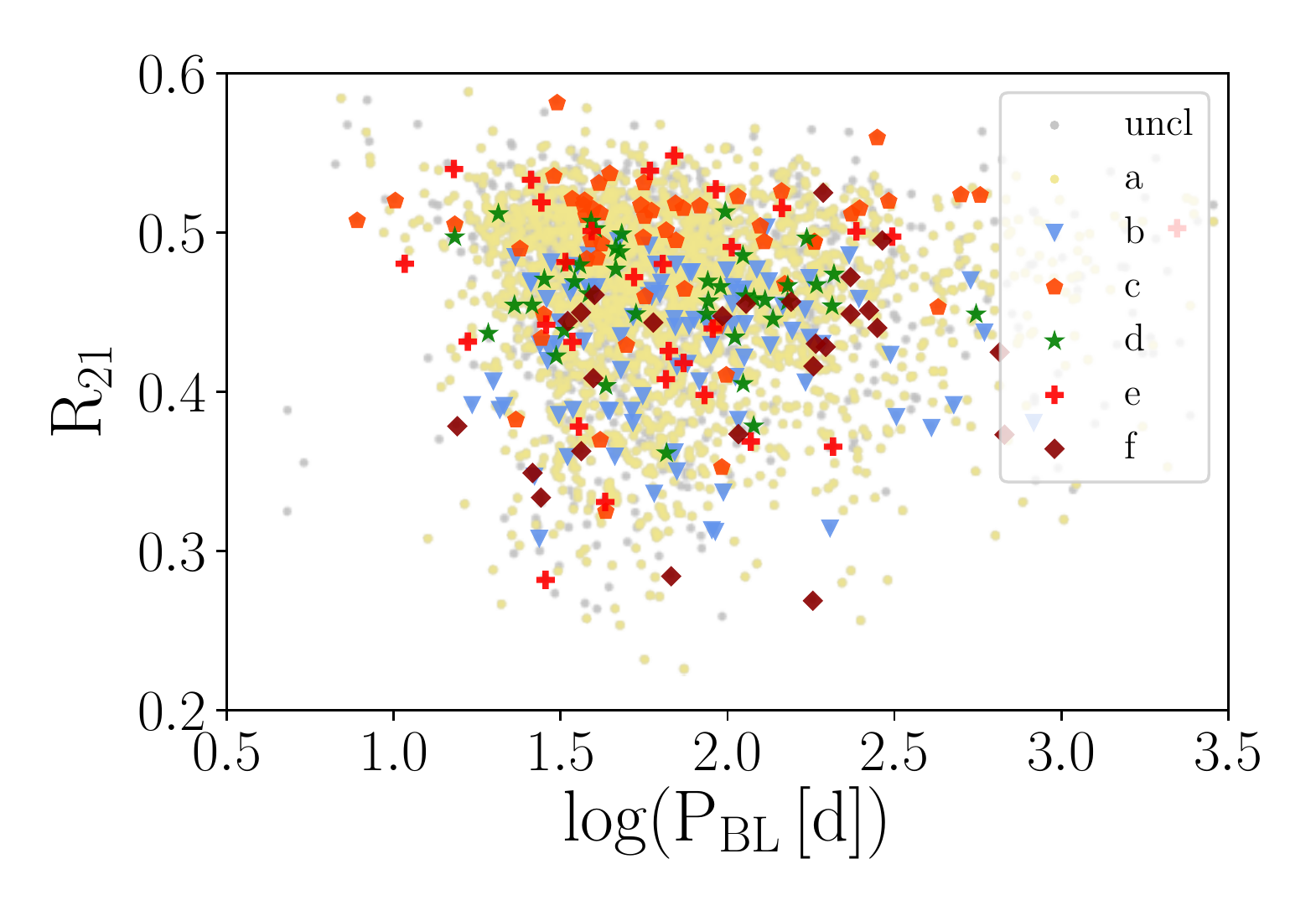}\includegraphics[width=.5\columnwidth]{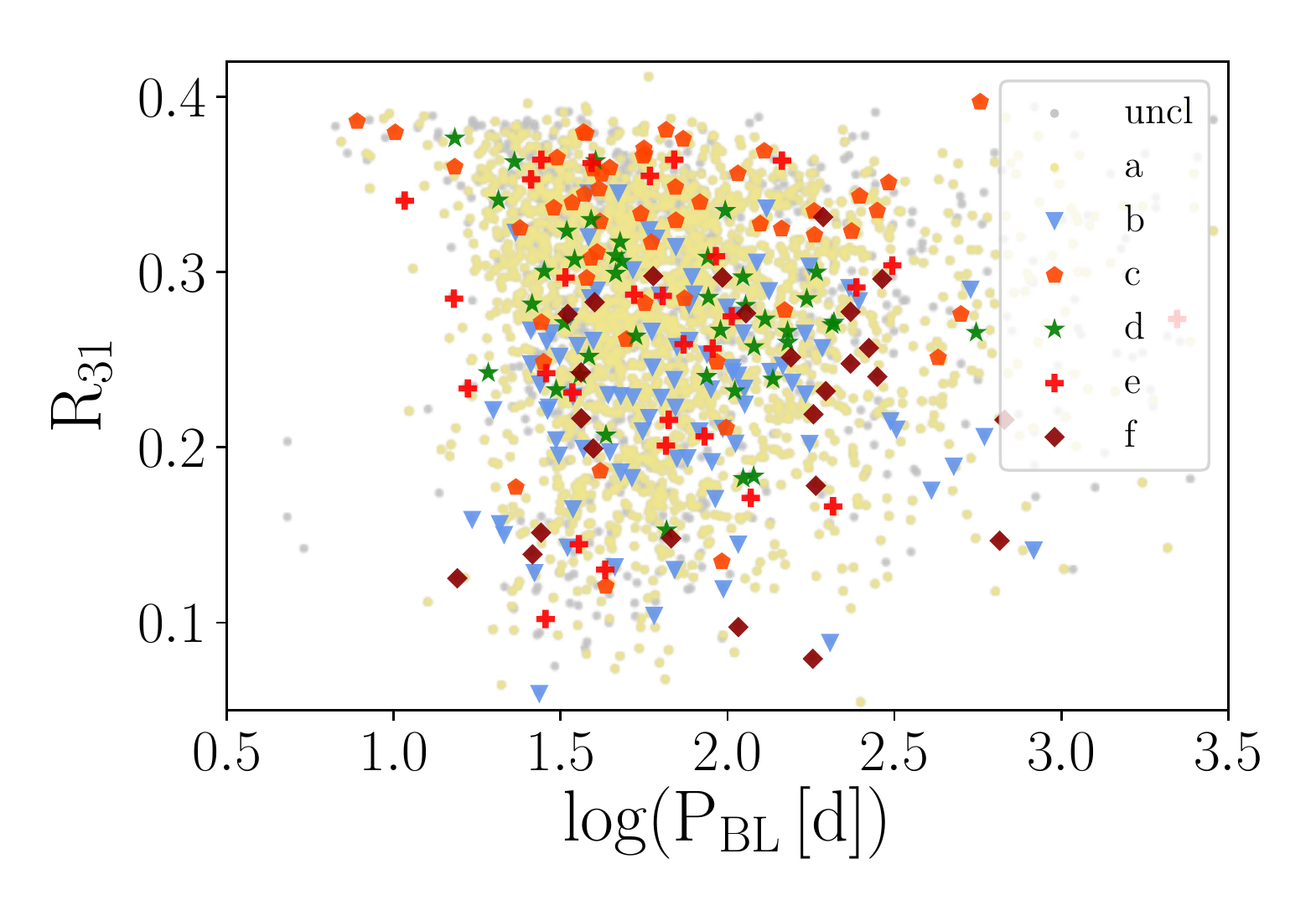}\\
	\includegraphics[width=.5\columnwidth]{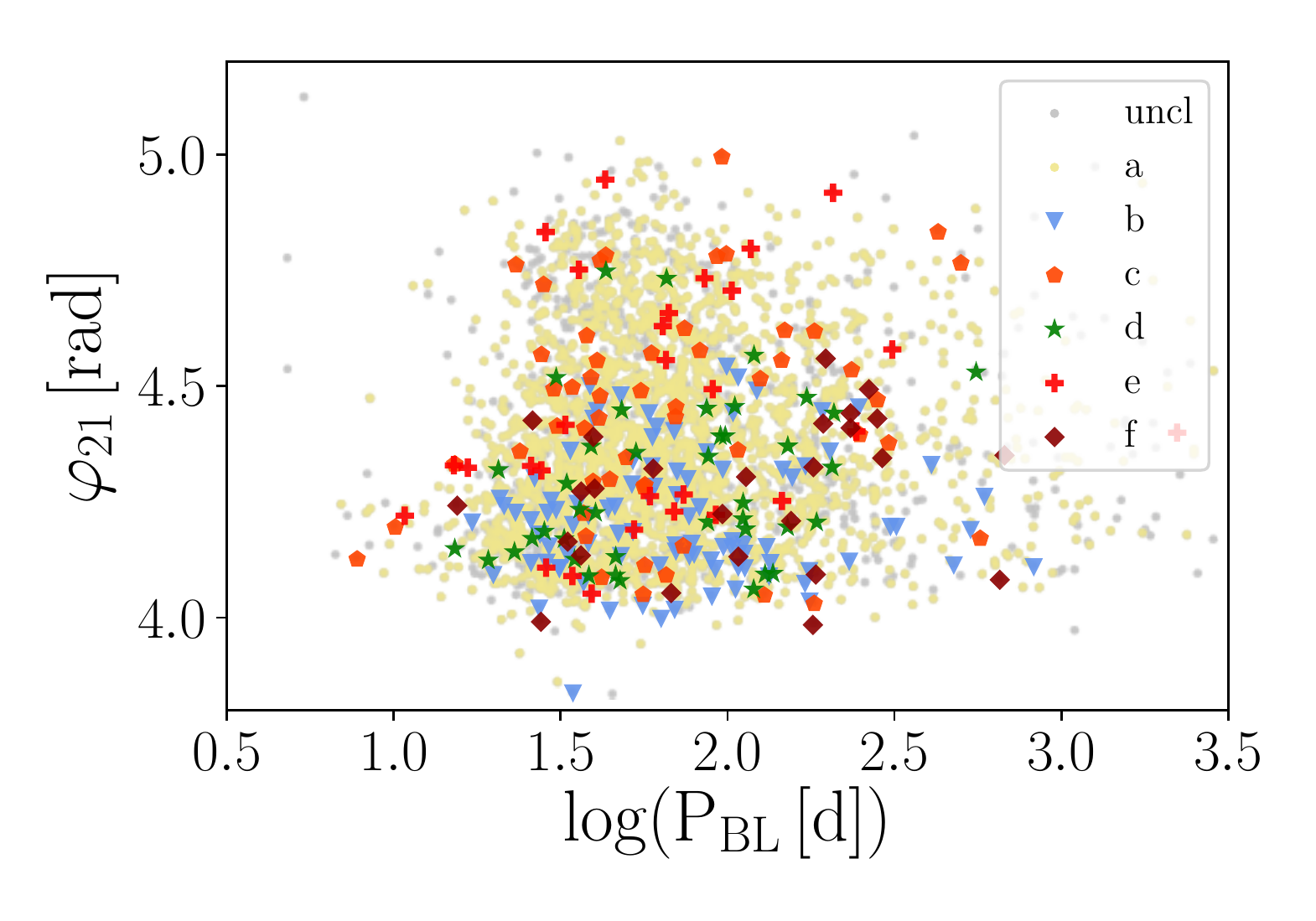}\includegraphics[width=.5\columnwidth]{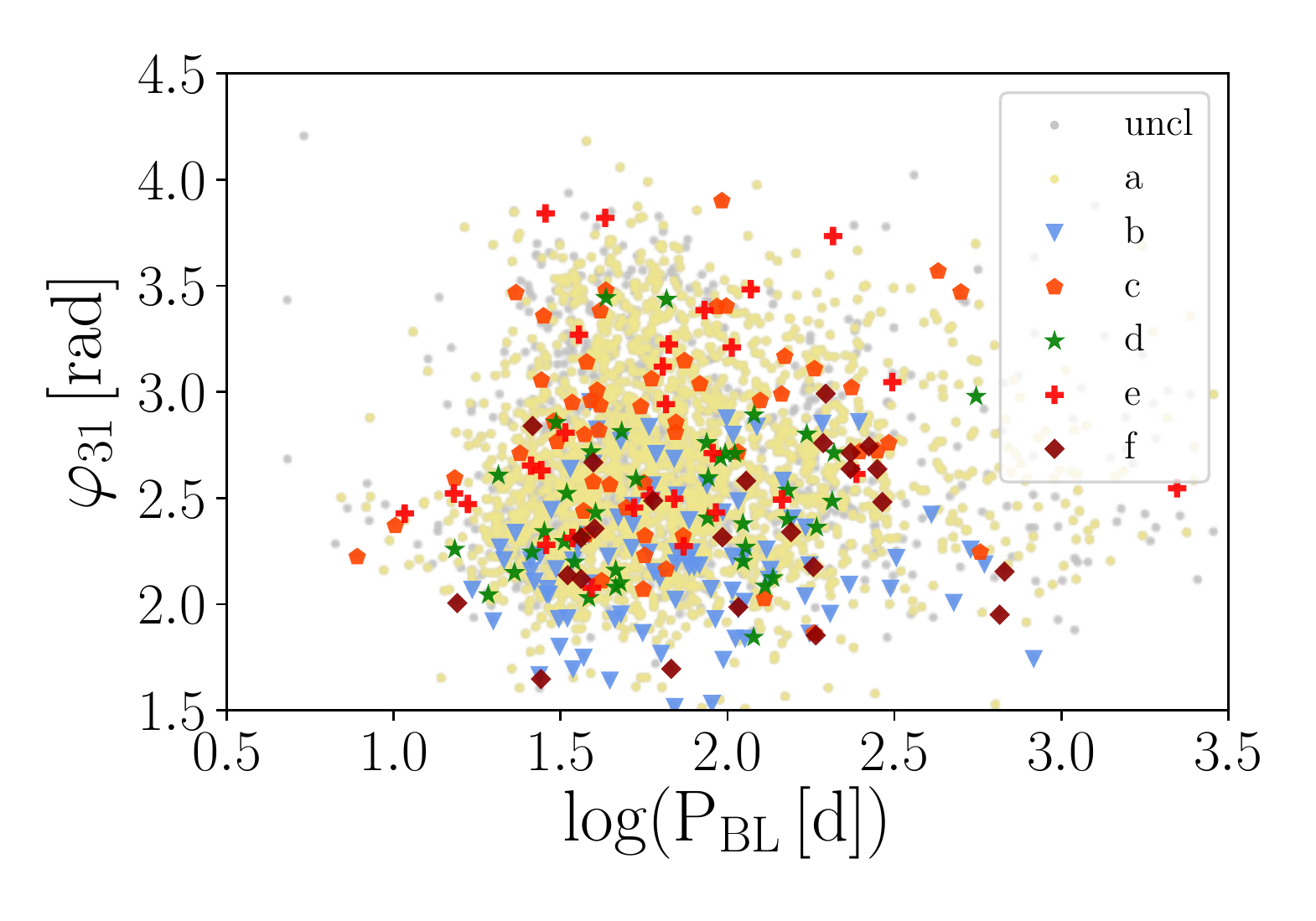}\\
    \caption{Low-degree Fourier coefficients of the sample stars. Amplitude coefficients $R_{21}$, $R_{31}$ ({\it top panels}) and phase coefficients $\varphi_{21}$ and $\varphi_{31}$ ({\it bottom panels}) as functios of the Blazhko period are shown.}
    \label{Fig:FourCoeff}
\end{figure}

We also investigated the connection between light-curve shape and the modulation properties. From the fit of the data with 10 pulsation harmonics we obtained amplitudes and phases of the particular components and calculated the Fourier amplitude and phase coefficients\footnote{The parameters introduced by \citet{simon1981} are defined as $R_{ij}=A_{i}/A_{j}$ and $\varphi_{ij}=\varphi_{i}-i\varphi_{j}$, where $A_{i,j}$ are amplitudes and $\varphi_{i,j}$ are phases of the particular harmonics.} describing the shape of the pulsation light curve (columns 23-26 in Table~\ref{Tab:Parameters}). As seen from Fig.~\ref{Fig:FourCoeff}, there is no apparent correlation between the low-degree Fourier coefficients and the length of the modulation period. In addition, no correlation between modulation amplitude or the {\it RT}$^{\rm TOP}_{\rm env}$ was observed.

\subsection{Photometric metallicity}\label{Sect:Metallicity}

\begin{figure}
\centering
\includegraphics[width=\columnwidth]{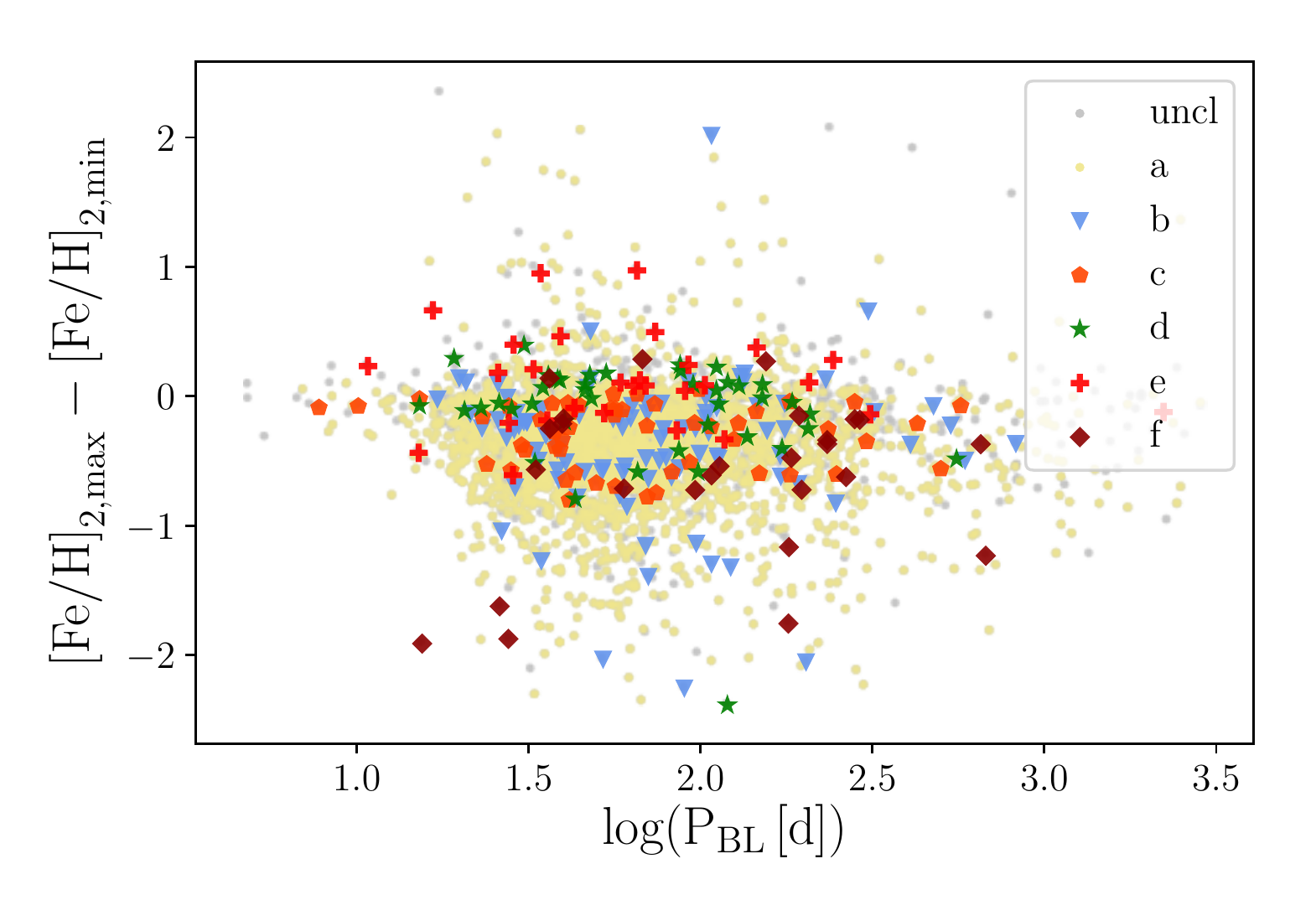}\\
\includegraphics[width=\columnwidth]{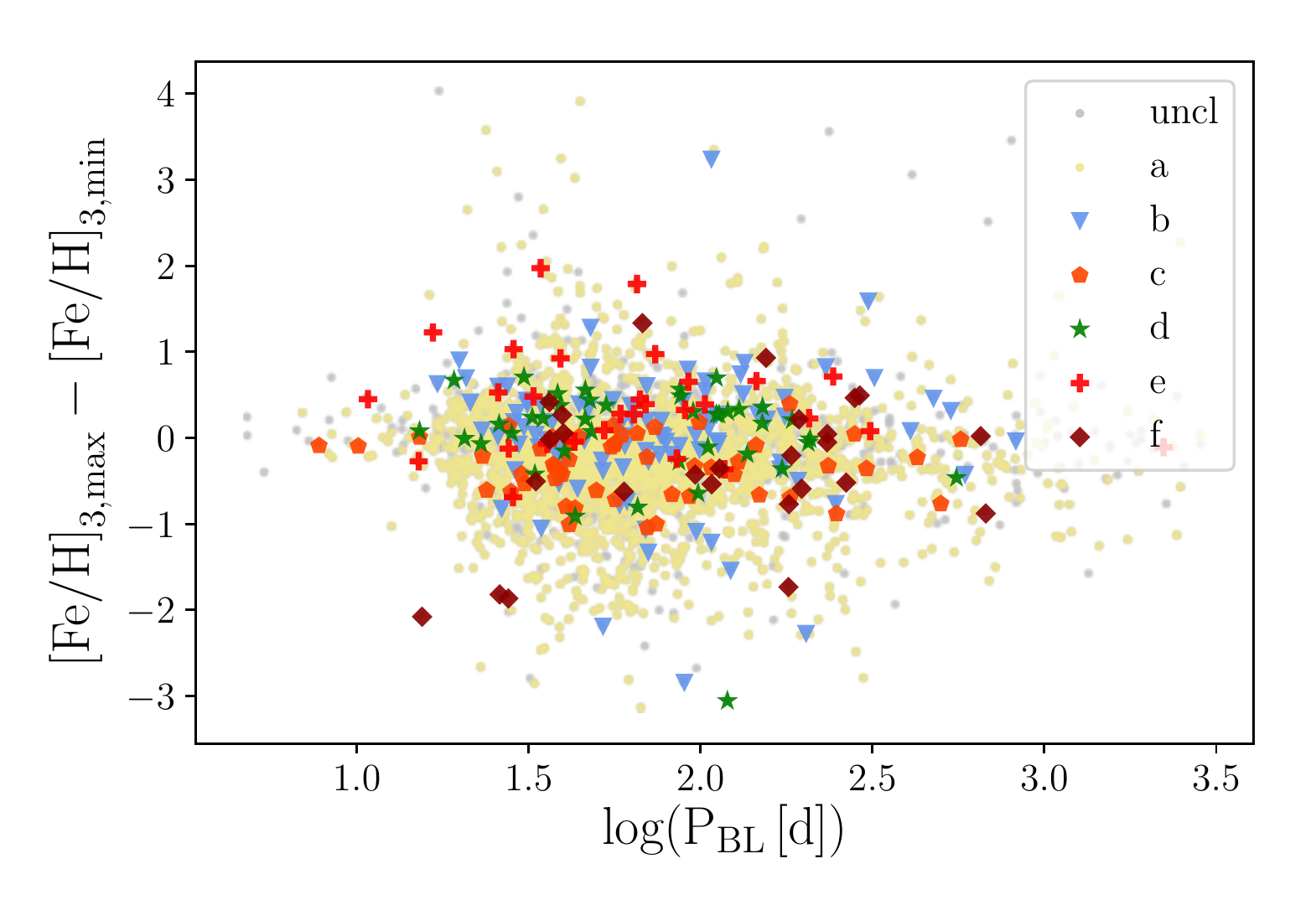}
\caption{Difference between photometric metallicity in maximum and minimum amplitude based on the two- and the three-parameter formula (top and bottom panel, respectively).}
\label{fig:Metallicity}
\end{figure}

The usability of the photometric metallicity among Blazhko stars has been discussed many times. We estimated the photometric metallicity using two- and three-parameter empirical relations from \citet{smolec2005}. Our results show that the difference of the metalicity determined at maximal and minimal Blazhko amplitude can be larger than 2-3\,dex (Fig.~\ref{fig:Metallicity}). The scatter is 0.5 and 0.7\,dex, respectively. Thus, the distance of a Blazhko star in the Galactic bulge ($\approx 8.3$\,kpc) determined in minimal and maximal Blazhko phases\footnote{Depending on the used formula from \citet{smolec2005} and the slope of {\it M}-[Fe/H] formula between 0.2 and 0.3 \citep{Clementini2003,smolec2005}.} differs by 400-800\,pc in average. Furthermore, the median metallicity of the whole sample based on the mean light curve slightly differs for two- and three-parameter relations ($-1.00$\,dex vs $-1.11$\,dex). This difference is due to the $A_2$ parameter in eq.~3 from \cite{smolec2005}. For comparison, \cite{Prudil2019OO} found that non-modulated fundamental mode RR~Lyrae stars in the Galactic bulge have median photometric metalicity $-1.05$\,dex using the three-parameter equation from \cite{smolec2005}. Therefore, since the majority of Blazhko stars in our sample lies at the short pulsation period end, one would expect that they would be more metal-rich using the eq.~3 in \cite{smolec2005}, but we find the opposite. The generally large difference shows that the full phase coverage over the whole Blazhko cycle is absolutely crucial \citep{jurcsik2009a,nemec2013} and the estimation of metallicity (and all other parameters calculated from metallicity) is generally not trustworthy in particular Blazhko stars and can be unreliable although in average the mean light curves give results similar to non-modulated stars \citep{Jurcsik2019}.

\subsection{The Blazhko stars with respect to the Bailey's diagram}

The open question of bimodality among the Milky Way globular clusters, based on the properties of RR~Lyrae stars they contain, also known as the Oosterhoff dichotomy \citep{oosterhoff1939}, is almost as old as the Blazhko effect itself. In this section, we investigate the relation between the Blazhko effect and Oosterhoff dichotomy. To complement this investigation, we utilized photometric data (mean magnitudes) and pulsation properties (amplitudes and pulsation periods) for RR~Lyrae stars in globular clusters M3 \citep{Benko2006M3dataIband,jurcsik2012a}, M5 \citep{Ferro2016M5data}, and M53 \citep{Dekany2009M53data,Ferro2012M53data}. We also used the identification of modulated stars from aforementioned papers. In the case of M53 we used Blazhko stars identified by \cite{Dekany2009M53data} which is based on period analysis of observed variables. We selected these globular clusters based on a high number of RR~Lyrae stars and information about their modulation.

\begin{figure}
\centering
\includegraphics[width=\columnwidth]{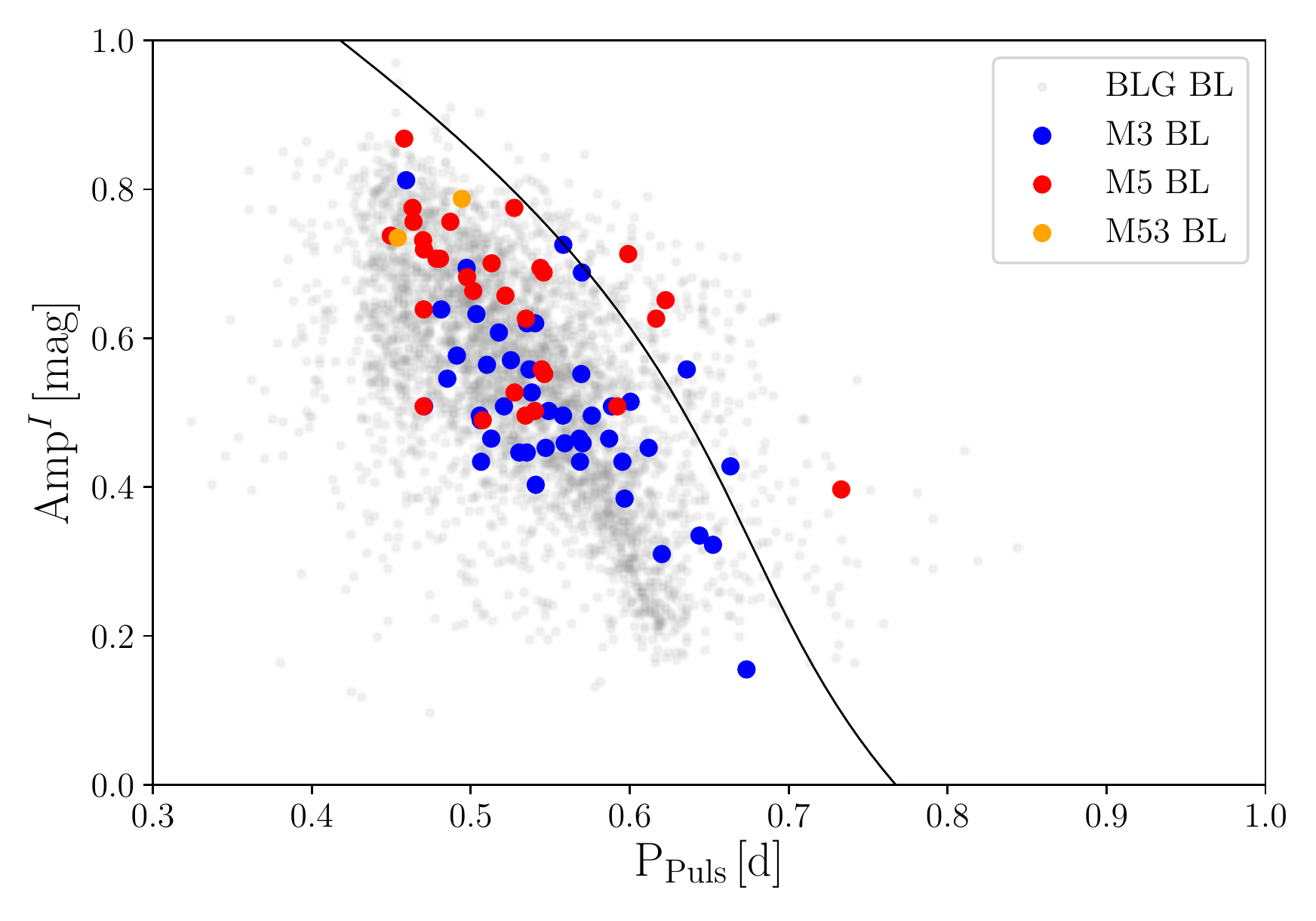}
\caption{The period-amplitude diagram for studied modulated stars. The light black dots represent RR Lyrae stars with the Blazhko effect from our sample, the blue, red, and orange points stand for modulated RRLs from globular clusters M3, M5 and M53, respectively. The black line separates variables into OoI and II.}
\label{fig:PeriodAmp}
\end{figure}

To divide stars into the Oosterhoff groups, we used the period-amplitude diagram in Fig.~\ref{fig:PeriodAmp}. For transformation of $V$-band amplitudes into $I$-band for RRLs in globular clusters with missing $I$ observations we used eq.~16 from \citet{Prudil2019OO}. The black line in Fig.~\ref{fig:PeriodAmp} outlines the boundary between Oosterhoff type I (hereafter OoI) and Oosterhoff type II (hereafter OoII) variables. This line is based on third-degree polynomial relation from table~2 in \citet{Prudil2019OO} and a shift in the pulsation period by 0.045\,day from \citet{Miceli2008}. We note, that the boundary between OoI and II globular clusters shifts with metallicity and the transformation of $V$ amplitudes into $I$ band can affect the association of individual modulated stars in the Oosterhoff groups. Neither of these effects can substantially change the overall distribution of modulated stars in the Oosterhoff groups, but in a handful of cases, it can change the association of a few RR Lyrae stars.

From Fig.~\ref{fig:PeriodAmp}, we clearly see that the vast majority of studied pulsators belong to the OoI group and that OoII group is deficient in modulated stars. In our sample, OoII stars constitute only 6.3\,per cent of all stars. Similar percentage of Blazhko RRLs is in globular clusters: M5 - 13.3\,per cent, and M53 - 0\,per cent of modulated stars belong to the OoII population. However, the data of M5 and M53 are not sufficient enough to define reliable BL percentage. The percentage of OoII BL stars in M3 is 40~per cent \citep{Jurcsik2017a}.

The small percentage of OoII BL stars when selected according to the their mean amplitudes is most-likely caused by the depreciation of the amplitudes among Blazhko stars in comparison with non-modulated variables. Using the amplitudes in the maximum of the Blazhko modulation phase (see Sec.~\ref{Sect:AmplitudeDetermination}) for sample variables and applying the aforementioned separation we get 22.6\,per cent representation of the Blazhko stars in the OoII group. This new ratio between OoI and OoII population of modulated stars corresponds better with the non-modulated sample used in \cite{Prudil2019OO} where they found 25~per cent of the Galactic bulge RR~Lyrae stars to be associated with the OoII group. The only morphological types that contain OoII stars are groups a and c. Other types do not contain OoII stars (see Table~\ref{Tab:MorphTypes}).

The average length of the modulation period is similar for both Oosterhoff groups. However, the amplitude of the modulation is of about factor two smaller for OoII RRLs than for the OoI stars. Similarly as for the OoI stars, the modulation amplitude of OoII BL stars gets smaller with increasing pulsation period.

\section{Discussion}\label{Sect:Discussion}

The variety of modulation-envelope shapes and amplitudes, modulation periods and the look of the frequency spectra is stunning. What drives the modulation? What rules which type of modulation is raised? Lots of the modulation properties can be easily deduced from the mathematical expressions. Modulation of a-type stars seem to be modulated with a simple sinusoid. 

The modulation shapes in b- and f-type stars (see Fig.~\ref{Fig:ModulationTypes}) can be explained with the non-sinusoidal modulation (see the examples in fig.~6 in \citep{benko2011}). The modulation function $f_{\rm m}^{\rm A}$ in Eq.~\ref{Eq:Modulation} can than be expanded in a sine series. The specific set of amplitudes and phases of the components of $f_{\rm m}^{\rm A}$ can produce very wide set of modulation envelopes. From that point of view, b- and f-type stars can actually be members of one single group. This idea is supported by the mixture and similar distribution of the modulation amplitudes which is most apparent from the Figures \ref{Fig:ModulationAmplitudes}, \ref{Fig:AblVsPbl} and \ref{Fig:AblVsPpuls} and Table~\ref{Tab:MorphTypes}. Also similar {\it RT$_{\rm env}$} of the upper envelopes supports the idea that groups b and f can actually constitute only one typological group. However, there must be some reason that these stars choose more often set of parameters that produce b-type stars rather than f-type stars. 

Types c, d, and e are not so easy to explain. It is very difficult to reproduce the flat bottom envelope (type c), double bottom envelope while the top envelope is single (type d) and larger bottom amplitude (type e) by simply employing amplitude modulation. Frequency modulation and/or other effects (double/multiple modulation, additional pulsation modes) must be present to obtain these features.  

Twenty-five per cent of the stars show additional modulation. It means that the modulation function $f_{\rm m}^{\rm A}$ (Eq.~\ref{Eq:Modulation}) has at least two components with different modulation frequencies and amplitudes. In presence of the double modulation, it is relatively commonly observed that the modulation frequencies are in resonance with small integer numbers \citep[e.g.][]{Sodor2011,benko2014a}. A nice example from our sample is OGLE-BLG-RRLYR-09745 showing 5:1 resonance between the two modulation components. By fine-tuning the parameters of the multiple-modulation components, it could be possible to explain the variation in the modulation cycles as well as the secular decrease/increase of the amplitude of the modulation \citep[see fig. 8 in][]{benko2014a}.  

The full description of the modulation in the large sample of stars including all the mentioned effects would require detailed investigation of every particular star or at least of several representatives which are typical for every of the six morphological types. The first results regarding the phase modulation of the Bulge RR Lyrae stars have been published recently by \citet{Jurcsik2020}. They found that the amplitude of the phase modulation decreases with increasing pulsation period similarly as the amplitude modulation studied in this paper. The full solution of the modulation in thousands of Galactic bulge RRLs would be the natural next step giving better picture of the Blazhko effect.

However, the description of the modulation is only the first step to understand the physics behind the modulation. The successful model has to explain all the variety of the observed features and what rules the choice of the modulation properties. Geometrical \citep[e.g.][]{Shibahashi2000} and resonant models \citep{vanHoolst2000} alone surely fail in explaining the variety of amplitudes, shapes and periods. The currently most promising 9:2 resonance model \citep{kollath2011} will have to comply in explaining this behaviour if it wants to succeed. However, what if we observe more effects simultaneously which are mixed in different ratios? Then the presence of non-radial pulsations as well as radial modes, aspect ratio and, alternatively, also some other physical effect, such as variable turbulent convection \citep{Stothers2006}, could be the important parameters. This is, however, pure guessing far out of our capabilities.

Out of the speculations, there must be a physical reason why stars which pulsate with longer periods may not have modulation periods under a certain limit (Fig.~\ref{Fig:BlazhkoPeriodDistribution}) and why the modulation amplitude decreases with increasing pulsation period (Fig.~\ref{Fig:AblVsPpuls}). This seems to be a common feature regardless the Oosterhoff groups. Stars which have longer periods are less dense (period of the fundamental mode is inversely proportional to the square root of the mean density). These stars are generally larger, cooler and more luminous than their short-period counterparts. Thus, the modulation amplitude and percentage of BL stars decreases towards the red edge of the instability strip and larger luminosities. The physical conditions apparently do not allow the modulation to rise in stars pulsating with long periods.

The fact that majority of modulated stars is located in the OoI group also tells a little bit about their physical properties in comparison with OoII stars. For example, the difference in masses between both Oosterhoff groups has been suggested in the past \citep{Catelan1992,Cacciari1993,Sandage2006,Prudil2019OO}, where the masses of OoII stars should be equal or larger in comparison with the OoI group. More importantly, the OoI variables lie close to the blue-edge of the instability strip and on are on average intrinsically hotter by several hundred Kelvins \citep{Prudil2019OO}. This would mean that modulated stars are on average hotter and possibly less massive than non-modulated stars, under the assumption that they are located on a zero-age horizontal giant branch. 

The reason for the existence of the two populations of modulation periods is a complete mystery (Fig.~\ref{Fig:BlazhkoPeriodDistribution} and \ref{Fig:DensityPlot}).

\section{Summary and conclusions}\label{Sect:Summary}
We investigated 3141 RRL stars from the Galactic Bulge using the OGLE-IV and III photometry. For all the studied stars, the Blazhko period could be reliably determined. The high-quality data and the large sample of modulated stars allowed us to investigate the modulation in a statistical matter. We determined the modulation periods, pulsation and modulation amplitudes, described the shape of the light curves and modulation envelopes. The visual inspection allowed us to identify six basic morphological shapes of the modulation and assign 2449 stars with good-enough data and large-enough amplitudes to one of these six modulation types (Fig.~\ref{Fig:ModulationTypes} and Table~\ref{Tab:MorphTypes}).

We also got information about the look of the frequency spectra - the position, amplitudes and asymmetry of the peaks. We also investigated the Oosterhoff phenomenon regarding the Blazhko effect. The full Table~\ref{Tab:Parameters} containing 30 columns is available online. We also provide plots showing the time distribution of the data, frequency spectra and phase light curves online (Fig.~\ref{Fig:FreqPhase}). 

Our main results and findings can be summarized in the following points:
\begin{enumerate}
    \item The distribution of the modulation periods (Fig.~\ref{Fig:BlazhkoPeriodDistribution}) and results from the Gaussian mixture modelling show that there could be two populations with different mean modulation periods of 48 and 186\,days (see Fig.~\ref{Fig:DensityPlot}). The pulsation periods of the stars from the two populations are similar and also the numbers of stars of different morphological types are similar in both populations. The only difference is that the incidence rate of the a-type stars is lower in the long-modulation period population. This result is in line with the asymmetric modulation envelopes, which are more common among stars with long Blazhko periods. The Blazhko valley, as we defined the area between the two populations in the \pb~vs \pp~plane, seems to be real.
    \item The only observed systematics in the length of the modulation period is that there is a bottom limit defining the shortest possible modulation period which increases with the pulsation period (bottom panel of Fig.~\ref{Fig:DensityPlot}). This is in accordance with \citet{jurcsik2005b}. We also observed a limit saying that in stars with longer modulation period the modulation amplitude can be larger (Fig.~\ref{Fig:AblVsPbl}).
    \item There is no simple dependence, nor a correlation between modulation period and amplitude of the modulation envelope as suggested by \citet{Benko2014b} observed in the OGLE RRLs we studied.
    \item The modulation amplitude decreases with increasing pulsation period (Fig.~\ref{Fig:AblVsPpuls}). 
    \item 70\,per cent of the stars have larger frequency peak at the right-hand side from the basic pulsation frequency suggesting that the phase difference $\phi_{\rm m}$ between amplitude and phase modulation is $\upi<\phi_{\rm m}<2\upi$.
    \item In 25\,per cent of stars we observed additional peaks suggesting additional modulation.
    \item About 60\,per cent of stars show asymmetric upper modulation envelope with steeper rise to maximum than the decrease to minimum. All stars of f-type behave like this.
    \item We warn in using photometric metallicity of the Blazhko stars because it can introduce uncertainty in the order of hundreds of parsecs in the distance of the Galactic bulge RR Lyrae Blazhko stars. Good sampling over the whole Blazhko cycle is undoubtely needed for a reliable metallicity estimation.
\end{enumerate}

Especially point (iv) suggests that there is some physical limit which prevent star to show the Blazhko effect. It seems that stars close to the red edge of the instability strip rarely show the Blazhko effect. This was already observed in a sample of 83 BL stars in M3 \citep{Jurcsik2019}. The lack of correlations between various parameters and the modulation period and amplitudes could mean that there is actually no simple general rule how and why some stars are modulated.

The next natural step of the investigation of the Blazhko effect will be thorough description of the modulation of the representatives of the different morphological types. Additional photometric and spectroscopic observations of the selected representatives will also help to set up the input parameters for the theoretical modelling.

\section*{Acknowledgements}
MS acknowledges the financial support of the Operational Program Research, Development and Education -- Project Postdoc@MUNI (No. CZ.02.2.69/0.0/0.0/16\_027/0008360). ZP acknowledges the support of the Hector Fellow Academy. JJ acknowledges the OTKA NN-129075 grant. We would like to thank the OGLE team for their great job with the observations of the GB.

\bibliographystyle{mnras}
\bibliography{references}

\section*{Appendix}\label{Sect:Appendix}

\begin{table*}
\caption{The example of the full table with all the important pulsation, modulation, frequency spectra and light curve parameters. All other parameters can be calculated from these parameters. The full table has 30 columns and 3141 rows and is available as a supporting material to the paper and at the CDS portal. The columns have the following meaning: {\it ID \#} -- five-digit number of the star complementing the full designation OGLE-BLG-RRLYR-XXXXX; \fz~-- the pulsation frequency; {\it Err}$_{f_{0}}$ -- the error of the pulsation frequency; {\it ARES}$_{f_{0}\pm0.3}$ -- the amplitude of the residuals around the $\pm0.3$\,c/d vicinity of $f_{0}$ after removing 10 pulsation components; \pb~-- the modulation period; {\it Err}$_{P_{\rm BL}}$ -- the error of the modulation period; $f_{-}$ -- the frequency of the left side peak; {\it Err}$_{f_{-}}$ -- the error of the frequency of the left side peak; $A_{-}$ -- amplitude of the left side peak in the frequency spectrum; $f_{+}$ -- the frequency of the right-hand side peak; {\it Err}$_{f_{-}}$ -- the error of the frequency of the right-hand side peak; $A_{+}$ -- amplitude of the right-hand side peak; $I_{\rm mean}$ -- the mean brightness of the star; $I_{\rm mean}^{\rm MAX}$ -- the mean brightness during the Blazhko maximum amplitude;  $I_{\rm mean}^{\rm MIN}$ -- the mean brightness during the Blazhko minimum amplitude; $A_{\rm mean}$ -- the mean amplitude of the light curve; $A_{\rm mean}^{\rm MIN}$ -- the amplitude of the light curve during the minimal-amplitude; $A_{\rm mean}^{\rm MAX}$ -- the amplitude of the light curve during the maximal-amplitude; $A_{\rm env}^{\rm TOP}$ -- the amplitude of the top envelope; $A_{\rm env}^{\rm BOT}$ -- the amplitude of the bottom envelope; {\it RT}$_{\rm env}^{\rm TOP}$ -- the rise time of the top envelope; {\it RT}$_{\rm env}^{\rm BOT}$ -- the rise time of the bottom envelope; $R_{21}$, $\varphi_{21}$, $R_{31}$, $\varphi_{31}$ -- amplitude and phase Fourier coefficients; {\it RT}$_{\rm LC}$ -- rise time of the mean light curve; Oo group -- Oosterhoff group; Morph type -- morphological type; Comments -- information about additional peaks (AdP), and period variation (UP).}
	\begin{tabular}{cccccccccc} 
		\hline
{\it ID} \#&\fz~(c/d)&{\it Err}$_{f_{0}}$~(c/d)&{\it ARES}$_{f_{0}\pm0.3} (mag)$&\pb~(d)&Err$_{P_{\rm BL}}$ (d)&$f_{-}$~(c/d)&{\it Err}$_{f_{-}}$~(c/d)&$A_{-}$ (mag)& $\rightarrow$ \\ \hline
00162&1.8304213&0.0000022&0.00142&171.23&1.02&1.82458&0.00003&0.0106&$\rightarrow$ \\
00172&2.0945254&0.0000021&0.00155&119.05&0.62&2.08610&0.00005&0.0094&$\rightarrow$ \\
00184&1.8750512&0.0000016&0.00136&47.24&0.08&1.85392&0.00008&0.0062&$\rightarrow$ \\
...&...&...&...&...&...&...&...&...&$\rightarrow$ \\
& & & & & & & & &  \\
& & & & & & & & &  \\ \hline
$\rightarrow$ &$f_{+}$&{\it Err}$_{f_{+}}$~(c/d)&$A_{+}$ (mag)&$I_{\rm mean}$ (mag)&$I_{\rm mean}^{\rm MAX}$ (mag)&$I_{\rm mean}^{\rm MIN}$ (mag)&$A_{\rm mean}$ (mag)&$A_{\rm mean}^{\rm MIN}$ (mag)&$\rightarrow$ \\ \hline
$\rightarrow$ &1.83613&0.00008&0.0049&15.941&15.943&15.948&0.656&0.528&$\rightarrow$ \\
$\rightarrow$ &2.10293&0.00004&0.0096&15.797&15.799&15.801&0.713&0.544&$\rightarrow$ \\
$\rightarrow$ &1.89622&0.00004&0.0111&15.769&15.781&15.757&0.730&0.668&$\rightarrow$ \\
$\rightarrow$ &...&...&...&...&...&...&...&...&$\rightarrow$ \\ \\
& & & & & & & & &  \\
& & & & & & & & &  \\ \hline
$\rightarrow$ &$A_{\rm mean}^{\rm Max}$ (mag)&$A_{\rm env}^{\rm TOP}$ (mag)&$A_{\rm env}^{\rm BOT}$ (mag)&{\it RT}$_{\rm env}^{\rm TOP}$&{\it RT}$_{\rm env}^{\rm BOT}$&$R_{21}$&$\varphi_{21}$ (rad)&$R_{31}$&$\rightarrow$ \\ \hline
$\rightarrow$ &0.7132&0.119&0.066&0.74&0.58&0.531&4.330&0.334&$\rightarrow$ \\
$\rightarrow$ &0.865&0.235&0.114&0.50&0.52&0.474&4.165&0.341&$\rightarrow$ \\
$\rightarrow$ &0.8304&0.073&0.037&0.44&0.50&0.485&4.162&0.367&$\rightarrow$ \\
$\rightarrow$ &...&...&...&...&...&...&...&...&$\rightarrow$ \\ \hline
& & & & & & & & &  \\
& & & & & & & & &  \\ \hline
$\rightarrow$ &  $\varphi_{31}$ (rad)&{\it RT}$_{\rm LC}$&Oo group&Morph type&Comments&&&&\\ \hline
$\rightarrow$ &2.619&0.17&1&a&--&&&&\\
$\rightarrow$ &2.295&0.15&1&a&--&&&&\\
$\rightarrow$ &2.296&0.14&1&--&--&&&&\\
$\rightarrow$ &...&...&...&...&...&&&&\\
 \hline
	\end{tabular}\label{Tab:Parameters}
\end{table*}

\bsp	
\label{lastpage}
\end{document}